\documentclass[12pt]{article}
\usepackage{epsfig}
\usepackage{citesort}
\usepackage{times}

\newcommand{\re}[1]{(\ref{#1})}
\newcommand{\eq}{\begin{equation}}
\newcommand{\eqx}{\end{equation}}
\newcommand{\eqn}{\begin{eqnarray}}
\newcommand{\eqnx}{\end{eqnarray}}

\newcommand{\nin}{\noindent}

\begin{document}
%
\begin{center}
{\Large \bf Femtosecond non-equilibrium dynamics of clusters
irradiated with short intense VUV pulses}\\
\vspace{10mm}

{\large B.~Ziaja $^{\ast,\dag}$ 
\footnote{Author to whom correspondence should be addressed. Electronic mail: ziaja@mail.desy.de}, 
H. Wabnitz $^{\ast}$, E. Weckert $^{\ast}$ 
and T. M\"oller $^{\ddag}$ 
}

\vspace{3mm}
$^{\ast}$ \it Hamburger Synchrotronstrahlungslabor HASYLAB,\\
Deutsches Elektronen-Synchrotron DESY,\\
Notkestr. 85, D-22603 Hamburg, Germany\\

\vspace{3mm}
$^{\dag}$ \it Department of Theoretical Physics,
 Institute of Nuclear Physics,
\it Radzikowskiego 152, 31-342 Cracow, Poland\\

\vspace{3mm}
$^{\ddag}$ \it Technische Universit\"at Berlin,\\
 \it Institut f\"ur Atomare Physik und Fachdidaktik,\\
 10623 Berlin, Hardenbergstrasse 36, Germany\\
 
\end{center}

\vspace{10mm}
\centerline{\Large November 2007}

\vspace{6mm}
\nin

\noindent
{\bf Abstract:}
The kinetic Boltzmann equation is used to model the non-equilibrium ionization phase that initiates the evolution of atomic clusters irradiated with single pulses of intense vacuum ultraviolet radiation. The duration of the pulses is $\leq 50$ fs and their intensity in the focus is $\leq 10^{14}$ W/cm$^2$. This statistical model includes various processes contributing to the sample dynamics at this particular radiation wavelength, and is computationally efficient also for large samples. Two effects are investigated in detail: the impact of the electron heating rate and the effect of the plasma environment on the overall ionization dynamics. Results on the maximal ion charge, the average ion charge and the average energy absorbed per atom estimated with this model are compared to the experimental data obtained at the free-electron-laser facility FLASH at DESY. Our analysis confirms that the dynamics within the irradiated samples is complex, and the total ionization rate is the resultant of various processes. In particular, within the theoretical framework defined in this model the high charge states as observed in experiment cannot be obtained with the standard heating rates derived with Coulomb atomic potentials. Such high charge states can be created with the enhanced heating rates derived with the effective atomic potentials. The modification of ionization potentials by plasma environment is found to have less effect on the ionization dynamics than the electron heating rate. We believe that our results are a step towards better understanding the dynamics within the samples irradiated with intense VUV radiation. 

\section{Introduction}


Unique properties of the short-wavelength free-electron-lasers (FELs) \cite{tesla,desy2006,slac,jap} emitting coherent radiation in ultraintense femtosecond pulses enable probing dynamic states  of matter, transitions and reactions happening within tens of femtoseconds, with wide-ranging implications to solid state physics, material sciences, and to  femtochemistry.  The focussed FEL beam is an excellent tool to generate and probe extreme states of matter \cite{xfelinfo2007,meyer1}. X-ray FELs (XFEL) will initiate novel structural studies of biological systems with single particle diffraction imaging. It is expected that single particle imaging will be applicable for the studies of the non-repetitive biological samples that cannot be performed with standard crystallographic methods \cite{l1,miao,gyula1,plasma4,liver1en,chapman}.

Rapid development of the research with FEL and the emerging experimental
results give strong motivation for theoretical studies of the ionization dynamics within the irradiated samples. Various processes are involved into this dynamics, and their contribution strongly depends on the radiation wavelength. Whereas the mechanisms of energy absorption and ionization within irradiated samples are well understood in case of irradiation with infrared radiation \cite{ditm,ditm1,jort,jort3,ishi}, this is not the case in the vacuum ultraviolet (VUV) regime. Electrons resulting from photoionizations of atoms with the intense VUV radiation form cold, strongly coupled electron plasma. The dynamics of these electrons is strongly influenced by their dense interacting surrounding. This effect is known as the plasma screening, and its contribution depends on charge densities and their temperatures. One of the consequences of the plasma screening is the modification of atomic potentials. It leads to lowering of the ionization potentials of ions and atoms, and also influences the cross-sections for interactions of charges within the plasma.

Full ab initio calculations of charge dynamics within strongly coupled plasmas are not available \cite{rostock2}. Therefore various approximate theoretical approaches are applied \cite{rostock,murillo}. Estimates of the plasma effects derived with these approximate models may differ significantly (e.g. screening models discussed in \cite{murillo}). Dedicated experiments could be helpful to sort out the relevant mechanisms. Among others, the data from the cluster experiments performed at the FLASH facility at DESY are available for theoretical analysis \cite{desy,desy2,desy1,desy3,desy4,desy7}. They cover the wavelength range from $100$ nm ($E_{\gamma}=12.7$ eV) down to $13$ nm ($E_{\gamma}=95.4$ eV). In this paper we will refer only to the first experiment, where xenon clusters were  irradiated with photons of energy, $E_{\gamma}=12.7$ eV. Pulse duration did not exceed $50$ fs, and the maximal pulse intensity was, $I\leq 10^{14}$ W/cm$^2$. Highly charged Xe ions (up to $+8$) of high kinetic energies were detected, indicating the strong energy absorption that could not be explained using the standard approaches \cite{desy5,desy2,desy7}. More specifically, the energy absorbed was almost an order of magnitude larger than that one predicted with classical absorption models, and the ion charge states were much higher than those observed during the irradiation of isolated atoms at the similar conditions. This indicates that at these radiation wavelengths some processes specific to many-body systems are responsible for the enhanced energy absorption. 

Several interesting theoretical models have been proposed in order to describe the evolution of clusters exposed to intense VUV pulses \cite{santra,santra1,siedschlag,georg,bauer,brabec,rusek}. Below we give a brief characteristics of some of them. Comprehensive review of the work performed until 2006 is given in Ref.\ \cite{rost1}. 
The physics underlying the dynamics within the irradiated clusters is complex. 
The first theoretical studies started with new ideas but introduced also some simplifications. In Refs. \cite{santra,santra1} the strong energy absorption within an irradiated atomic cluster resulted from the enhanced inverse bremsstrahlung (IB) heating of quasi-free electrons. This rate was estimated with an effective atomic potential \cite{joa} which represents the attraction of the nucleus and the average screening effect of bound electrons surrounding the nucleus. Therefore the distribution of bound electronic charge around the nucleus is smooth. An energetic electron that passes through the inner of an atom/ion is then scattered by an effective positive charge, $Z_{eff}$, larger than the net charge of the ion. This effect leads to the enhancement of the total IB rate that is proportional to the squared charge of the scatterer. This mechanism was first explored in Ref.\  \cite{santra}. It lead to the production of high charges within the irradiated clusters. These high charges were created in a sequence of electron impact ionizations. Relative distributions of ion charges were similar to those observed in the experiment \cite{desy}. However, this first study considered the ionization within an infinitely extended homogeneous cluster, and was not taking into account the dynamics of charges. The IB rate was calculated perturbatively. Also, impact ionization was treated approximately
with a simplified rethermalization scheme.

This model was improved by the same group in Ref.\ \cite{santra1}. A model of cluster expansion was added. IB rate was recalculated with the Debye-screened Herman-Skillman potential, using a non-perturbative approach. Recombination and impact ionization processes were treated explicitly. Simulations performed with this improved model again showed the formation of highly charged ions within the irradiated clusters.

We stress here that the derivation of the IB rate with the effective atomic potential as performed in Ref.\ \cite{santra,santra1} is in contrast to the standard approaches that assume Coulomb potentials of point-like ions \cite{krain1,krain,rozmus}. The heating mechanism similar as in Ref. \cite{santra} was recently successfully tested in Ref. \cite{deiss}. It was applied to model the heating of quasi-free electrons in large rare-gas clusters irradiated with infrared laser pulses. These electrons were heated during elastic large-angle backscatterings on ion cores. Potentials of ions were modelled with the parametrized atomic potential similar to that one in Ref. \cite{santra}. Absolute x-ray yield obtained with this effective atomic potential was in better agreement with the experimental data than that one obtained with the bare Coulomb atomic potential.

A different mechanism of the strong energy absorption within an irradiated cluster was proposed in Ref.\ \cite{siedschlag,georg}. According to this model, high charges within small clusters can be created in a sequence of single photoionization processes. Collisional ionizations via electron impact and recombinations are neglected. For isolated Xe atoms and ions only single photoionizations: $Xe + \gamma \rightarrow Xe^{+}$ occur. This is due to the low energies of the incoming photons, $E_{\gamma}=12.7$ eV, that slightly exceed the ionization potential of a neutral Xe atom, $E_{+1}=12.1$ eV. Within a cluster, atomic potentials overlap at the interatomic distances small enough. Lowered interatomic potential barriers are then formed. These barriers are further suppressed with the increasing ion charge \cite{jort3,siedschlag}, facilitating the inner ionization of bound electrons into the cluster. At the potential barriers low enough further photoionizations are possible. Higher charge states can then be formed. 

The electrons released during the photoionization processes are confined within the cluster (inner ionization). They are heated with the IB process enhanced by the presence of highly charged ions. The effective heating rate obtained with
point-like ions is similar to that of Refs. \cite{krain,krain1}. When the electrons are hot enough, they start to escape from the sample. This initiates its expansion. 

Molecular dynamics simulations were performed in order to test this model.  Distribution of ion charges obtained for xenon cluster consisting of $80$ atoms was in a good agreement with experimental data. The model has not been tested for larger clusters ($N_{at}\geq 200$) yet. The non-homogeneous distribution of charges within the cluster (consisting of positively charged outer shell and a neutral core) predicted with these simulations is confirmed by the recent experimental findings for the mixed cluster systems \cite{desy6}.

Another heating process, alternative to IB, was proposed in Ref.\ \cite{brabec}. This many-body process called the many-body-recombination may occur within dense strongly coupled electron-ion systems. Electrons are heated in a sequence of recombination and photoionization events. They collide with atoms and ions, creating higher charges via impact ionizations. Ions of charge up to $+7$ were predicted with this model for the $Xe_{80}$ cluster.

Among other models of laser-cluster interaction we mention a quasi-classical model of Bauer \cite{bauer} and the Thomas-Fermi calculations \cite{rusek}. 
Results obtained with these models followed qualitatively the experimental findings.

So far the models describing the interaction of the rare-gas clusters with the intense VUV radiation were characterized. A model for the absorption of VUV photons in metals and warm-dense-matter was proposed in \cite{meyer2,meyer3,meyer4,krenz}. The basis for this model was the microscopic theory of IB that used the IB rates calculated by Krainov \cite{krain1,krain} for slow and fast electrons. The predictions obtained with this model were in a good agreement with the data from the transmission experiments. An interesting mechanism of femtosecond switching from transmission to reflection within irradiated Al foils was identified with the simulations performed in Ref.\ \cite{meyer2}. This ultrafast switching was due to the coincidence between the VUV radiation frequency and the plasma frequency.


\section{Motivation for this study}


As we have shown above, various theoretical models have been developed in order to explore the strong absorption and the presence of the high charge states observed in the first VUV experiment. However, we can expect that if all enhancement factors proposed with these models would be included within one model, it would probably lead to the absorption rates much higher than those experimentally observed. 

With this theoretical study we aim to test the influence of two effects: i) 
the impact of the IB heating rate, ii) the impact of the modification of the ionization potentials (due to the plasma environment) on the non-equilibrium ionization dynamics within the large Xe clusters ($N=2500$ atoms) irradiated with a flat pulse of intense VUV radiation. Parameters of the pulse are: photon energy $E_{\gamma}=12.7$ eV, intensity $10^{12}-10^{14}$ W/cm$^2$ and duration $<50$ fs. We will consider two different IB rates: i) that one calculated by Krainov for point-like ions in Refs.\ \cite{krain1,krain}, ii) the enhanced IB rate proposed by Santra in Ref. \cite{santra}. In order to estimate the effect of the plasma screening and the charged ion environment, we will treat two limiting cases: i) the case when atomic energy levels are shifted due to the plasma effects, and ii) the case when no energy level shifts are assumed. Atomic potentials then correspond to the potentials of isolated atoms/ions. 

In order to follow the cluster evolution, we will use the statistical Boltzmann approach proposed in \cite{ziajab,ziajab1}. Our Boltzmann code solves the full kinetic equations for electron and ion densities within the irradiated sample. Particles (represented as particle densities) interact with the mean electromagnetic field created by all charges and also with the laser field. The microscopic interactions: photoabsorptions, collisional processes (also IB) enter these equations as rates. These rates are included into the two-body collision terms, and are estimated either from experimental data or with theoretical models. 

Below we write a general form of kinetic equations within an irradiated sample. The coupled semi-classical Boltzmann equations for single electron, $\rho^{(e)}({\bf r,v},t)$, and ion densities, $\rho^{(i)}({\bf r,v},t)$, where $i=0,1,\ldots, N_J$ denotes the ion charge, and $N_J$ is the maximal ion charge are:

{\small
\eq
\partial_t\rho^{(e)}({\bf r,v},t)+{\bf v}\cdot\partial_{\bf r}\rho^{(e)}({\bf r,v},t) 
+{e\over m}\left( {\bf E}({\bf r},t)+{\bf v}\times {\bf B}({\bf r},t) \right)
\cdot\partial_{{\bf
v}}\rho^{(e)}({\bf r,v},t) =
\Omega^{(e)}(\rho^{(e)},\rho^{(i)},{\bf r,v},t),
\label{be}
\eqx
\eq
\partial_t\rho^{(i)}({\bf r,v},t)+{\bf v}\cdot\partial_{\bf r}\rho^{(i)}({\bf r,v},t) 
-{i e\over M}\left( {\bf E}({\bf r},t)+{\bf v}\times {\bf B}({\bf r},t) \right)
\cdot\partial_{{\bf v}}
\rho^{(i)}({\bf r,v},t) =
\Omega^{(i)}(\rho^{(e)},\rho^{(i)},{\bf r,v},t).
\label{bi}
\eqx
}

\noindent
These equations include the total electromagnetic force acting on ions and electrons. Collision terms, $\Omega^{(e,i)}$, describe the changes of the electron/ion densities with time. These changes are due to short-range microscopic  processes. Type of processes involved in the sample dynamics depends on the radiation wavelength.

Our simulation tool follows the non-equilibrium femtosecond dynamics within spherically symmetric samples of large or moderate size. As it evolves the particle densities, the computational costs does not scale directly with the number of atoms within the sample. During the sample evolution no assumption of local thermodynamic equilibrium (LTE) is made \cite{lee}. Therefore this code can be applied to describe dynamics of samples irradiated with ultra-short pulses of the duration 
less than a few femtoseconds. i.e. less or comparable with the thermalization timescale. Techniques for generating such ultra-short pulses have been already 
discussed in \cite{saldin1,saldin2,saldin3,saldin4}.
The non-equilibrium treatment of sample evolution is an advantage when comparing our programme to the hydrodynamic codes. These codes are efficient for large samples but they include simplifying assumptions on the dynamics of charges such as LTE condition or the collective movement of charges. If the thermalization timescales are short comparing to the pulse length, hydrodynamic models are reliable tools to follow the evolution of irradiated samples. However, at shorter pulses sample evolution should be treated with  non-equilibrium models.

Comparing with the state from Refs. \cite{ziajab,ziajab1}, our model has been significantly extended and improved. More interactions are now treated and included into the programme. We will discuss them in the next section.


\section{Evolution of samples exposed to intense VUV radiation}



\subsection{Interactions}


We will now specify the physical processes that have been included into our model of the charge dynamics within the irradiated cluster:

\noindent
{\bf 1.} { \it Photoionizations, collisional ionizations and
elastic scatterings of electrons on atoms/ions.} As in \cite{ziajab}, the cross sections for these interactions were estimated with the experimental data on atomic cross sections.

\noindent
{\bf 2.} {\it Long-range Coulomb interactions of charges.} Interactions with external laser field are treated within the dipole approximation. This approach is justified by the small spatial size of the irradiated cluster of a radius $\sim 36$ \AA, when compared to the wavelength of laser radiation ($\sim 100$ nm). Following our estimates from Ref. \cite{ziajab}, we expect that the
attenuation of the laser beam is small, and we neglect it. Interactions of a charge with internal field are modelled as electrostatic interaction of this charge with the mean field created by all charges. This mean field is estimated
with the densities of positive and negative charges.

\noindent
{\bf 3.} {\it Heating of electrons due to the inverse bremsstrahlung process (IB).} The heating rate is estimated either: i) with the Krainov heating rates calculated for slow and fast electrons \\ \cite{krain,krain1} or ii) with the quantum mechanical cross-section obtained with the Born approximation \cite{kroll}, using the effective atomic potential proposed in Ref.\ \cite{santra}.

\noindent
{\bf 4.} {\it Modification of atomic potentials by electron screening 
and ion environment.} In order to calculate the energy level shifts due to the electron screening we use the hybrid potential proposed in \cite{murillo}. This potential was constructed to match the ion-sphere picture (limit of strongly coupled plasma) at small distances and Debye-H\"uckel picture (limit
of weakly charged plasma) at large distances. Therefore it can adapt to the changing conditions during the evolution of an irradiated cluster. This
hybrid potential extends the standard treatment proposed by Stewart and Pyatt \cite{stewart}, as it may additionally account for degeneracy effects. 
Modification of the ionization cross sections due to the plasma effects was estimated as in Ref. \cite{rostock} by including the shifted ionization potentials into these cross sections. This is the first order approximation that may underestimate the magnitude of the cross sections, if the energy level shifts are large \cite{rostock1}.
 
Following the ideas proposed in Ref. \cite{jort3,siedschlag,georg}, we considered also the influence of the charged environment of an ion within the plasma on the  ionization potential of this ion. We included an estimate of the ionization potential shift due to the lowering of the interatomic potential barriers. As the quasi-free electrons within the cluster screen the ion charge, this shift was calculated from the overlap of the screened (Debye-H\"uckel) potentials of the neighbouring ions. If the screening by electrons is efficient,
as e.\ g.\ in the interior of the cluster, the interatomic potential barriers obtained with the screened potentials will be higher than those estimated in Ref.\ \cite{jort3,siedschlag} with bare Coulomb potentials. As a result, the reduction of the ionization potentials due to the barrier suppression will be smaller than this obtained with the bare Coulomb potentials. In case of low electron screening, e.\ g.\ at the surface layer of the cluster, the estimated shifts should approach those obtained with bare Coulomb potentials.  
 
\noindent
{\bf 5.} {\it Shielded electron-electron interactions.} They induce fast thermalization of the sample. The respective Fokker-Planck term \cite{shark} representing this interaction was added to the right-hand-side of the Boltzmann 
equation for the electron density.

The following processes were neglected within our model: excitation and deexcitation of bound electrons, multiionization processes, $Xe^{+q}+e/\gamma\rightarrow Xe^{+(q+n)}$ at $n>1$, and ionization by internal electric field at the cluster edge. 

We estimated the contribution of multiphoton processes in detail. Using the cross sections for multiphoton ionization of Xe ions, calculated in Ref.\ \cite{santra2}, we estimated the two-photon and three-photon ionization rates at $I=10^{14}$ W/cm$^2$ and $E_{\gamma}=12.7$ eV. They were: $R_{2\gamma}=\sigma_2 (I/E_{\gamma})^2=1.1$ fs$^{-1}$ for the $Xe^+\rightarrow Xe^{+2}$ process, and $R_{3\gamma}=\sigma_3 (I/E_{\gamma})^3=0.02$ fs$^{-1}$ for the $Xe^{+2}\rightarrow Xe^{+3}$ process accordingly. For comparison, the average collisional ionization rate for the process $Xe^+\rightarrow Xe^{+2}$, estimated after $\sim2$ fs of the exposure, when all atoms are singly ionized, was $R_e \sim 5$ fs$^{-1}$. This implies that the multiphoton process $Xe^+\rightarrow Xe^{+2}$ can contribute only early in the exposure, and at later times the collisional ionization dominates. Therefore the contribution of the multiphoton processes to the total ionization rate is of a minor importance. The other multiphoton processes, $Xe^{+q} \rightarrow  Xe^{+(q+1)}$, where $q>1$, have even lower rates, and are therefore negligible within the cluster environment.

We also estimated the three-body recombination rate using the Zeldovich-Raizer formula for singly charged plasma in LTE \cite{plasma3}. This formula was derived, assuming the detailed balance principle. It was estimated as $\sim 1$ fs$^{-1}$ early in the exposure (low electron temperatures) and $\leq 0.04$ fs$^{-1}$ later in the exposure (high electron temperatures). Higher ion charges may lead to the enhancement of these recombination rates \cite{hahn}. As the simulation times are less than $100$ fs, the recombination processes are omitted within the present model.

Additional pressure on ions due to the recoil effects during electron-ion collisions was neglected due to large mass difference between electrons and ions, $m_e/M_{Xe} \sim 10^{-5}$ and the short simulation timescales.  

At present our simulation follows the ionization phase of the cluster evolution. The expansion phase will be treated in the forthcoming papers (see Appendix).


\subsection{Ionization dynamics modelled with Boltzmann solver}


We will now demonstrate the ionization dynamics within an irradiated cluster on the following example. We will study the evolution of the $Xe_{2500}$ cluster exposed to a flat FEL pulse of intensity, $I=6\cdot 10^{13}$ W/cm$^2$ and the duration of $\Delta t=10$ fs. Interactions listed in the preceding subsection are included in this simulation. The IB process is modelled with the enhanced IB rate from Ref.\ \cite{santra}. Atomic potentials correspond to 
those ones of the isolated atoms.

We define the integrated charge densities, $n_j(v,t), n_j(r,t)$, that will further be used to analyze the simulation predictions:
\begin{eqnarray}
n_j(v,t)&\equiv&\int\,\rho^{(j)}({\bf r},{\bf v},t)\,\,d^3r,
\nonumber\\
n_j(r,t)&\equiv&\int\,\rho^{(j)}({\bf r},{\bf v},t)\,\,d^3v.
\label{dens}
\end{eqnarray}
The densities, $\rho^{(j)}({\bf r},{\bf v},t)$, are charge densities in phase-space. Indices are: $j=e$ for the electron density, and $j=0,\ldots,N_J$ for the ion densities. These densities are evolved with Eqs.\ \re{be}, \re{bi}. Integrated densities are then obtained with Eqs.\ \re{dens}.
  
Our Boltzmann solver solves Eqs.\ \re{be}, \re{bi} in phase-space within the simulation box of a finite size. The limits are: $0 < r < 120$ \AA$\,$  and $0 < v < 140$ \AA/fs. The number of grid points is correspondingly $60$ in real space and $140$ in velocity space.

\begin{figure}
\vspace*{0.5cm}
\centerline{a)\epsfig{width=7cm, file=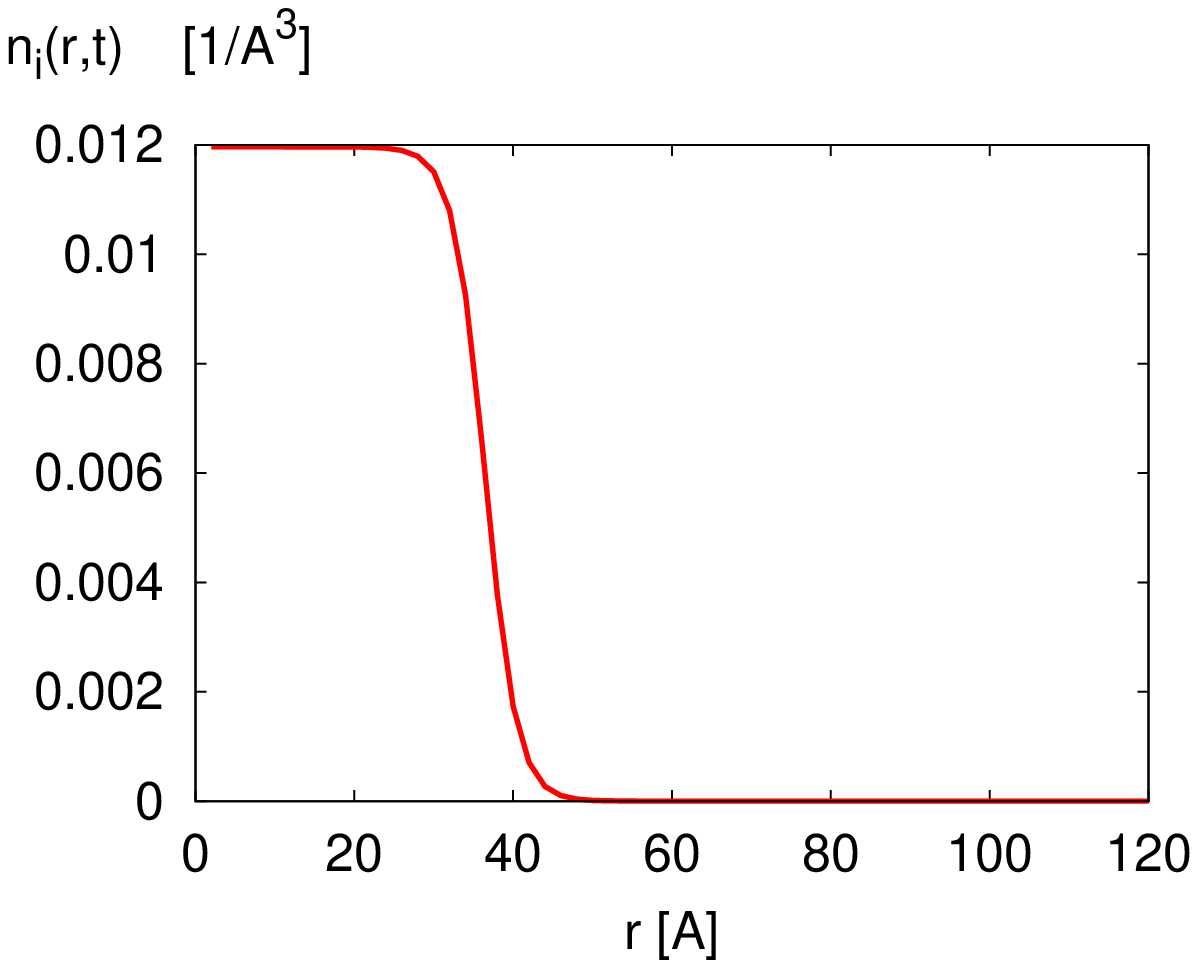}
b)\epsfig{width=7cm, file=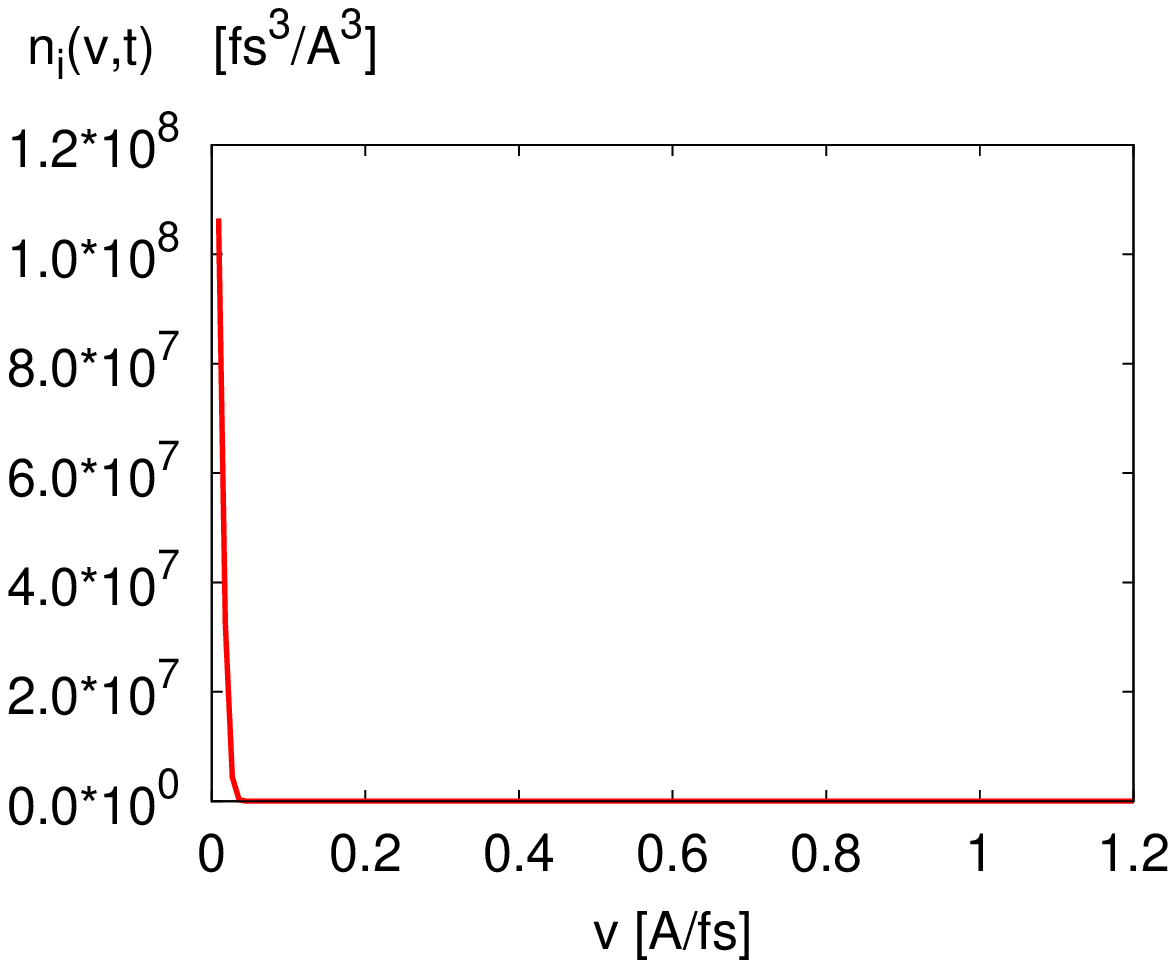} }
\caption{Initial configuration at $t=0$ fs: a) atomic density as a function of $r$, b) atomic density as a function of $v$.}
\label{init}
\end{figure}

The initial configuration is given by a smooth uniform density function representing a spherically symmetric cluster consisting of $2500$ neutral xenon atoms (Fig.\ \ref{init}a). Here the edges of the sample are smoothed in order to facilitate computation. The density in the center corresponds to that of the xenon cluster. The radius of this cluster is $\sim 36$ \AA. The initial velocity distribution of atoms is given by a delta function, $\delta(v)$, approximated by the narrow Gaussian distribution function (Fig.\ \ref{init}b). In order to check how our results are influenced by the choice of the width of this Gaussian distribution we performed a test simulation at the ten-times smaller width. The results obtained agreed with the previous ones. This confirms that our results are not biased by this specific parametrization of the initial ion velocity distribution at the considered simulation timescales. 

We can distinguish two main phases of the sample evolution: {\bf ionization phase} and {\bf expansion phase} (not discussed here in detail). The non-equilibrium {\bf ionization phase} starts after the sample is exposed to the laser radiation and lasts until the saturation of ionizations is reached. Its duration depends on the pulse length and the pulse intensity. Here this phase may last up to several tens of femtoseconds. The maximal pulse length considered is, $\Delta t=50$ fs, and the pulse intensity lies in the range, $I \sim 10^{12}-10^{14}$ W/cm$^2$.

\noindent
{\bf Photoionization:}\\
Ionization phase starts with single photoionizations that release single electrons from the outer shell of Xe atoms: $\gamma(E_{\gamma})+Xe \rightarrow e(E_e) + Xe^{+1}$, where $E_{\gamma}$ denotes the energy of the incoming photon, and $E_e$ is the energy of the released photoelectron. In this case, the photoelectron energy will be $\sim 0.6$ eV, as the ionization threshold for Xe is $E_{+1}=12.1$ eV. Early in the exposure only single photoionizations are possible, due to the low energies of the incoming photons. When the density of emitted electrons grows, plasma effects lead to the lowering of the ionization potentials within the sample. If these energy shifts are sufficiently large, further ionizations of Xe ions via single photon absorption can occur. The photoabsorption process: $\gamma +Xe^{+1}\rightarrow Xe^{+2}$ is treated within our model.

\begin{figure}
\vspace*{0.5cm}
\centerline{a)\epsfig{width=7cm, file=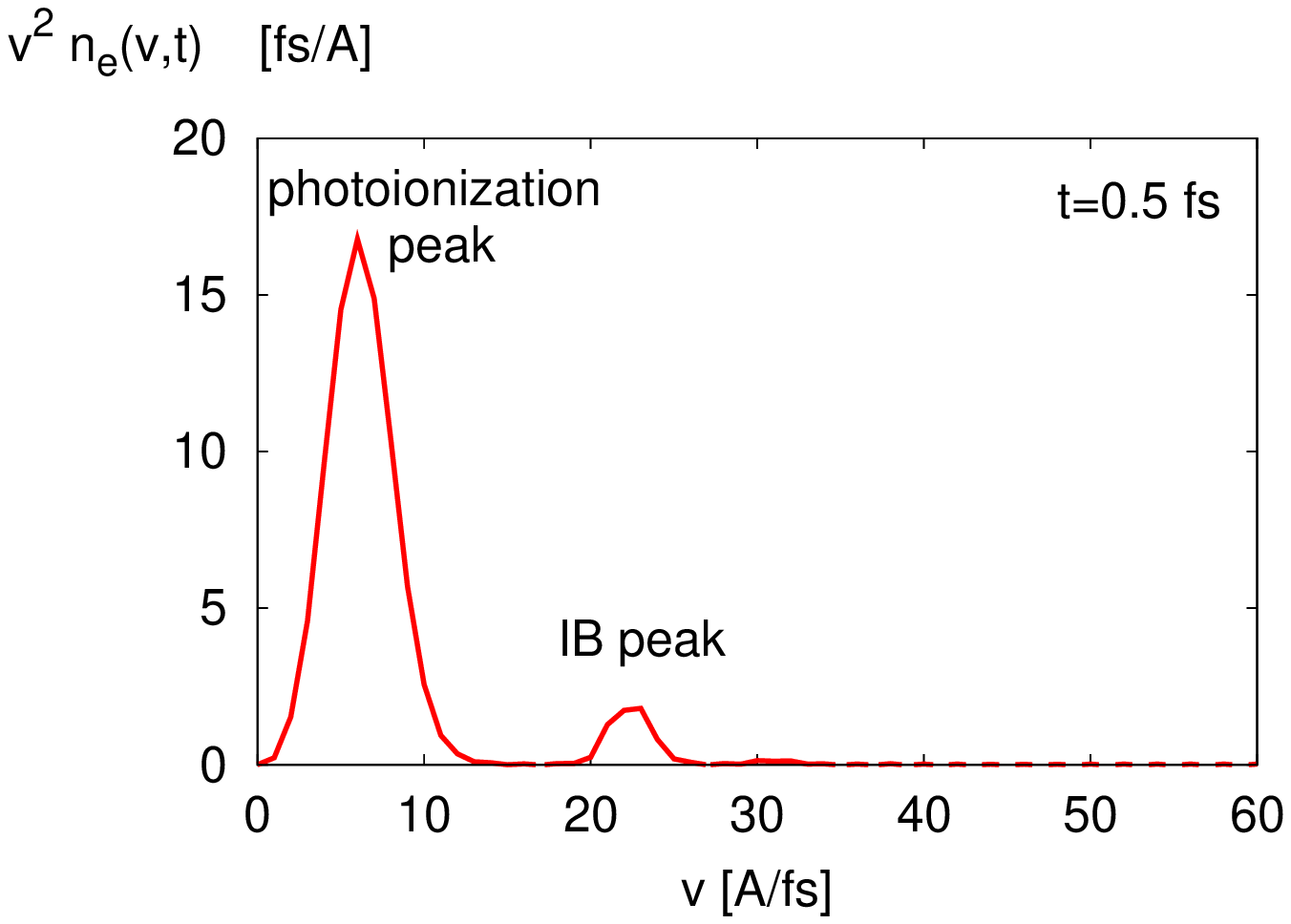}
b)\epsfig{width=7cm, file=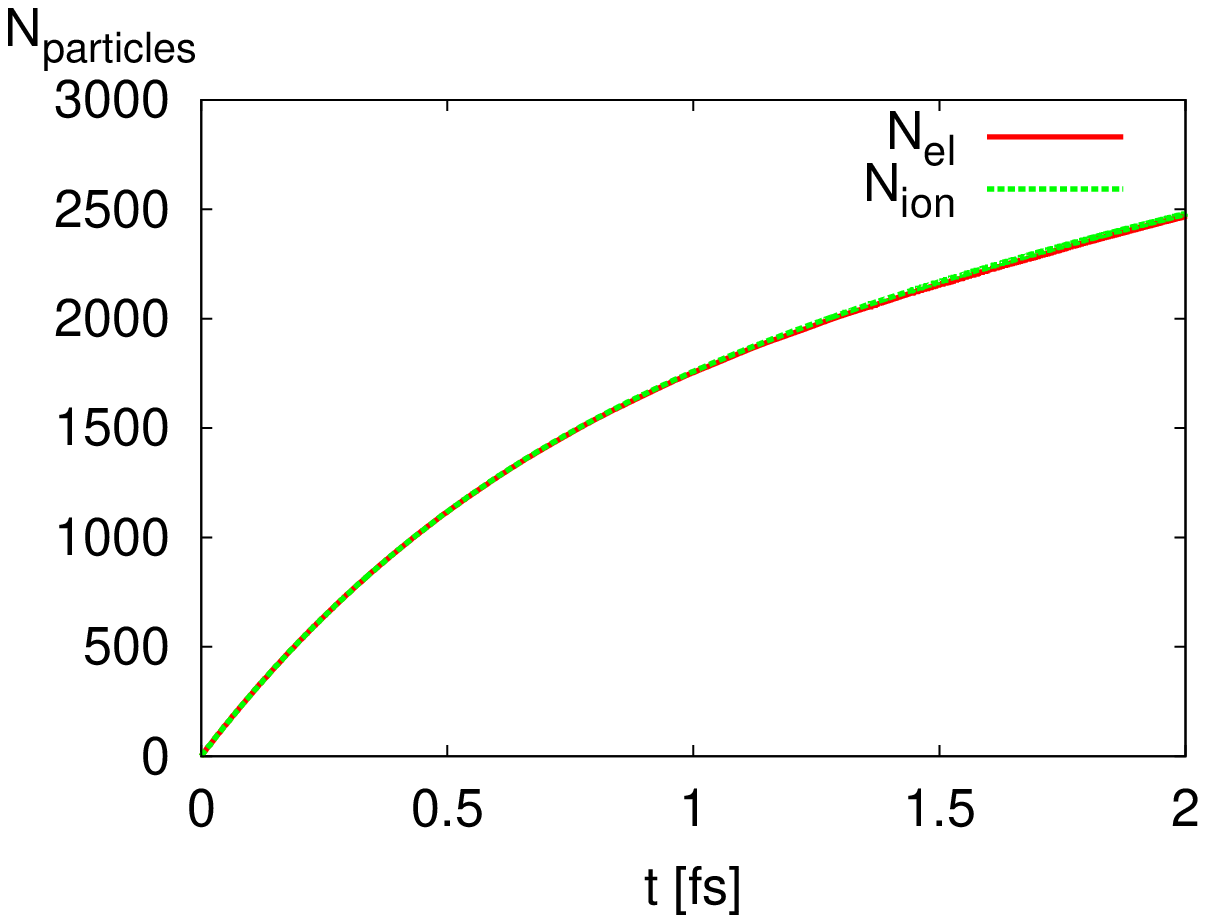} }
\caption{Photoionization phase: a) strong photoionization peak at the electron velocity distribution at $t=0.5$ fs, b) number of electrons released and gross-number of ions, $N_{ion}=\sum_{i=1}^{N_J}\,i\cdot N_i$, created as a function of time. Up to $\sim 2$ fs of the exposure the electron population is dominated by photoelectrons.}
\label{phot}
\end{figure}

\begin{figure}
\vspace*{0.5cm}
\centerline{a)\epsfig{width=5cm, file=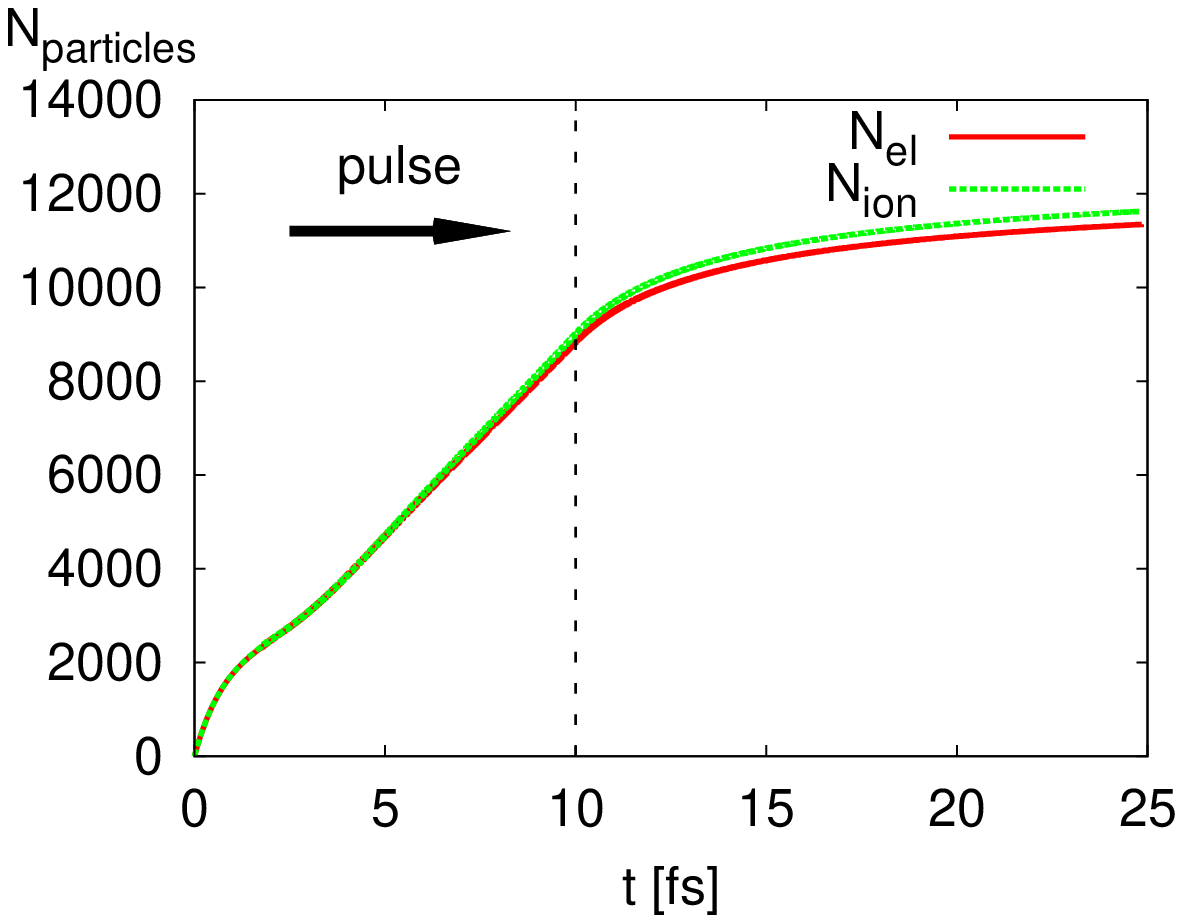}
b)\epsfig{width=5cm, file=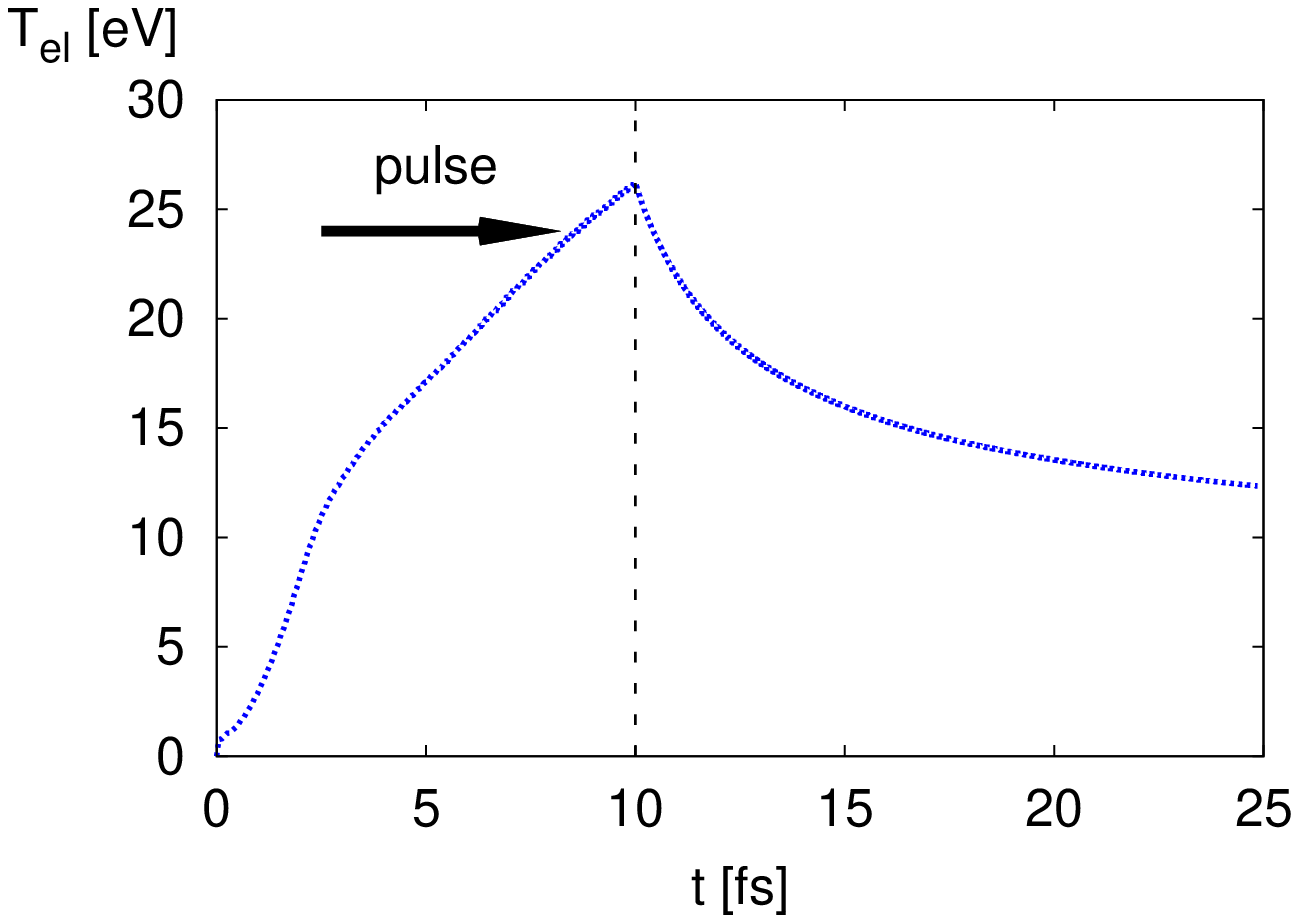}
c)\epsfig{width=5cm, file=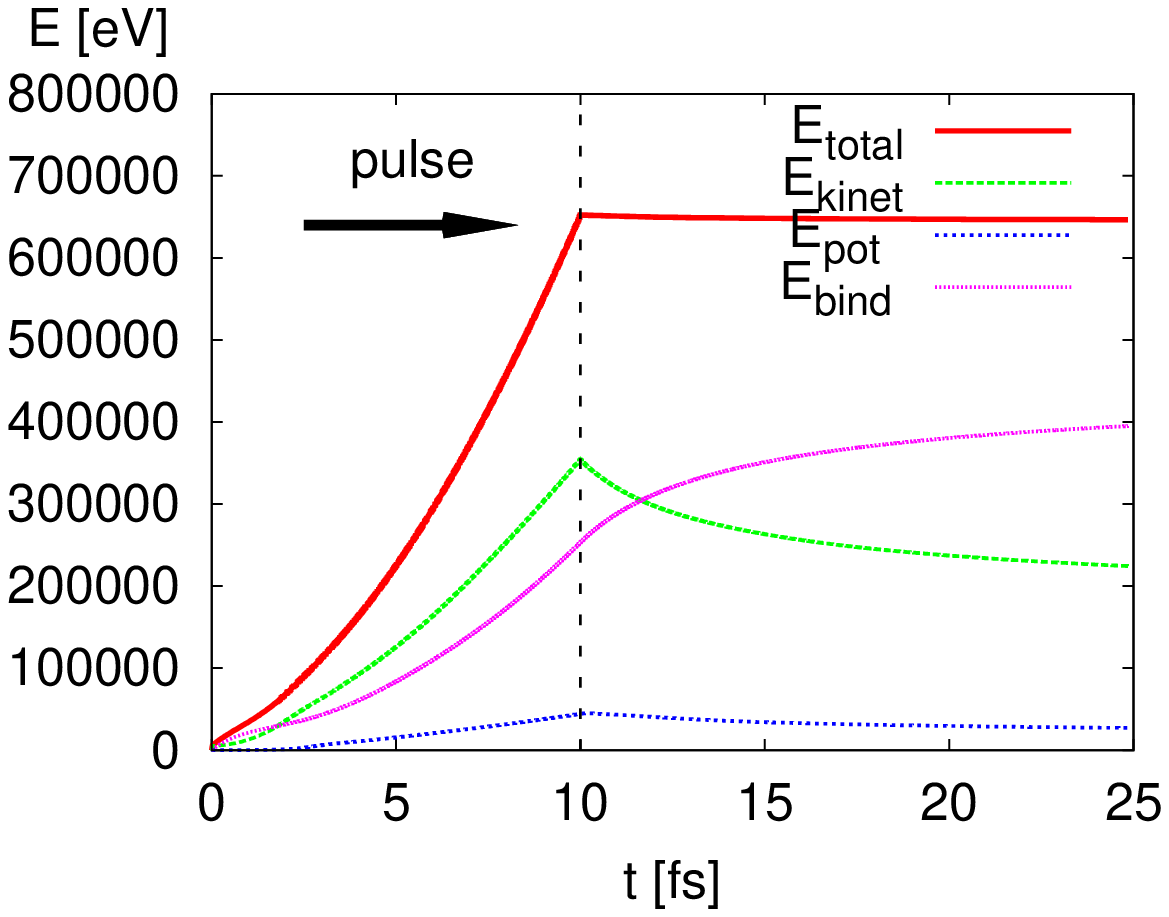}}
\caption{Global parameters within the irradiated cluster as a function 
of time: a) total number of electrons, $N_{el}$, and gross number of ions,
$N_{ion}$, b) temperature of electrons, $T_{el}$, and c) total energy within the sample, $E_{total}$: $E_{total}=E_{kinet}+E_{pot}+E_{bind}$, is the sum of the kinetic energies of electrons and ions, $E_{kinet}$, the potential energy within the electron-ion system, $E_{pot}$, and the total energy that was needed to release electrons from atoms and ions during the ionization processes, $E_{bind}$.}
\label{global}
\end{figure}

Fig.\ \ref{phot}a shows the photoionization peak in the electron velocity distribution (on the left) after $\sim 0.5$ fs. The second peak (on the right) corresponds to the single photoabsorption during IB process. It is clearly visible that photoionization remains the dominating process of the electron release until $\sim 2$ fs, whereupon the number of released electron starts to saturate (Fig.\ \ref{phot}b). With the estimates of the screening effects included into this simulation, if the electrons could not gain more energy through a heating process, the number of ionizations would saturate shortly after this time. As we included an enhanced heating rate into the model, after $\sim 2$ fs electrons are hot enough to initiate collisional ionizations, and we observe a fast linear growth of the charge numbers due to these processes (Fig.\ \ref{global}a). Number, $N_{ion}$, denotes the gross number of ions, $N_{ion}=\sum_{i=1}^{N_J}\,i\cdot N_i$, where $N_i$ is the number of ions of charge, $i$, and $N_J$ denotes the maximal ion charge.

\begin{figure}
\vspace*{0.5cm}
\centerline{a)\epsfig{width=7cm, file=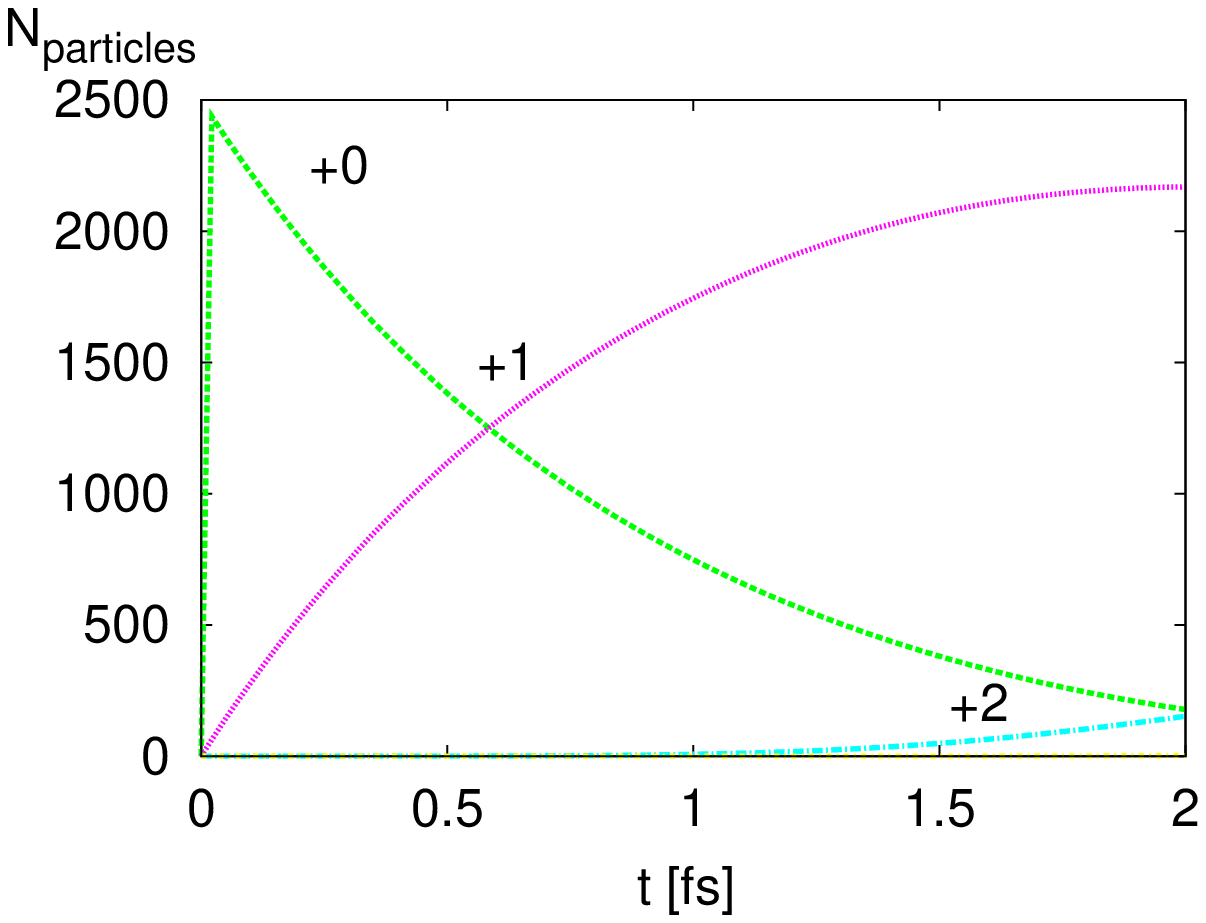}
b)\epsfig{width=7cm, file=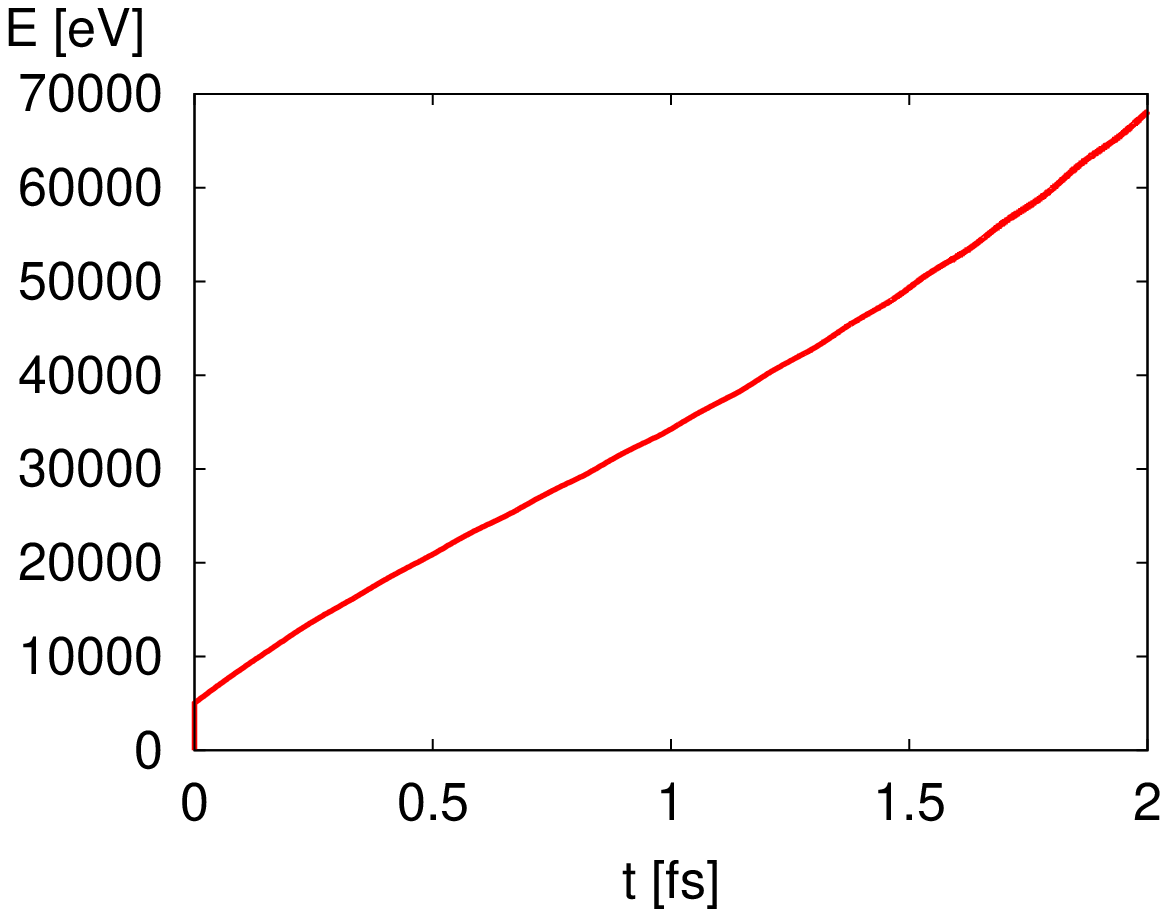} }
\caption{Energy absorption during photoionization: 
a) production of single Xe ions from Xe atoms at $t\leq 2$ fs, b) total energy absorbed by the sample as a function of time. Up to 
$\sim 1.5$ fs total energy absorption is a linear function of time.
At later times, it becomes non-linear due to the inverse bremsstrahlung process.}
\label{photabs}
\end{figure}

Energy absorption for the single photoionization process is described 
by a formula: ${dE_{abs}}/{dt}= I\,\sigma_{\gamma}\,N_{at}(t)$, where $N_{at}(t)$ is the number of the neutral atoms at time $t$, $N_0$ is the initial number of atoms, $I$ denotes the pulse intensity and 
$\sigma_{\gamma}=\sigma_{\gamma}(E_{\gamma})$ is the total photoionization cross section. The number of neutral atoms decreases exponentially, $N_{at}(t)=N_0\cdot e^{-I\sigma_{\gamma}t/E_{\gamma} }$, as it can be seen in Fig.\ \ref{photabs}a. This formula implies that the energy absorbed during photoionizations, $E_{abs}$, will change linearly with $t$ at small $t$ (Fig.\ \ref{photabs}b).

\noindent
{\bf Heating through IB}\\
When ion and electron densities are large enough, heating or cooling of electrons with the inverse bremsstrahlung process starts: $e(E_e)\pm n\gamma \rightarrow e(E_e\pm n\hbar \omega)$. Inverse bremsstrahlung is defined as a stimulated emission or absorption of radiation quanta by a free electron in the field of an ion. Within the approximation of Ref.\ \cite{kroll}, if the field strength parameter, $s={ {e{\bf E}_0}\over {m \omega} }{ {1}\over {\hbar \omega} }$ is low, and the free electrons are slow, single-photon exchanges dominate.
As $s>1$, or as the free electrons are fast and may undergo collisions with ions with large momentum transfers, multi-photon exchanges can occur. This latter (limiting) case can be identified with the classical impact picture \cite{kroll}.

During the heating the total energy absorption within the sample can be non-linear with respect to the exposure time and the pulse intensity: 
${{dE_{abs}}/{dt}} \propto N_{ion}(I,t)\,\sigma_{IB}(I)\,N_{el}(I,t)$,
as the total numbers of ions and electrons, $N_{ion}(I,t)$ and $N_{el}(I,t)$, change with the pulse intensity and the exposure time. Due to these non-linearities, we may expect the different amount of radiation energy absorbed within the sample at the same
radiation energy flux, $F=\int dt\,I(t)=const$, but at various pulse
intensity shapes, $I(t)$. 

\begin{figure}
\vspace*{0.5cm}
\centerline{a)\epsfig{width=7cm, file=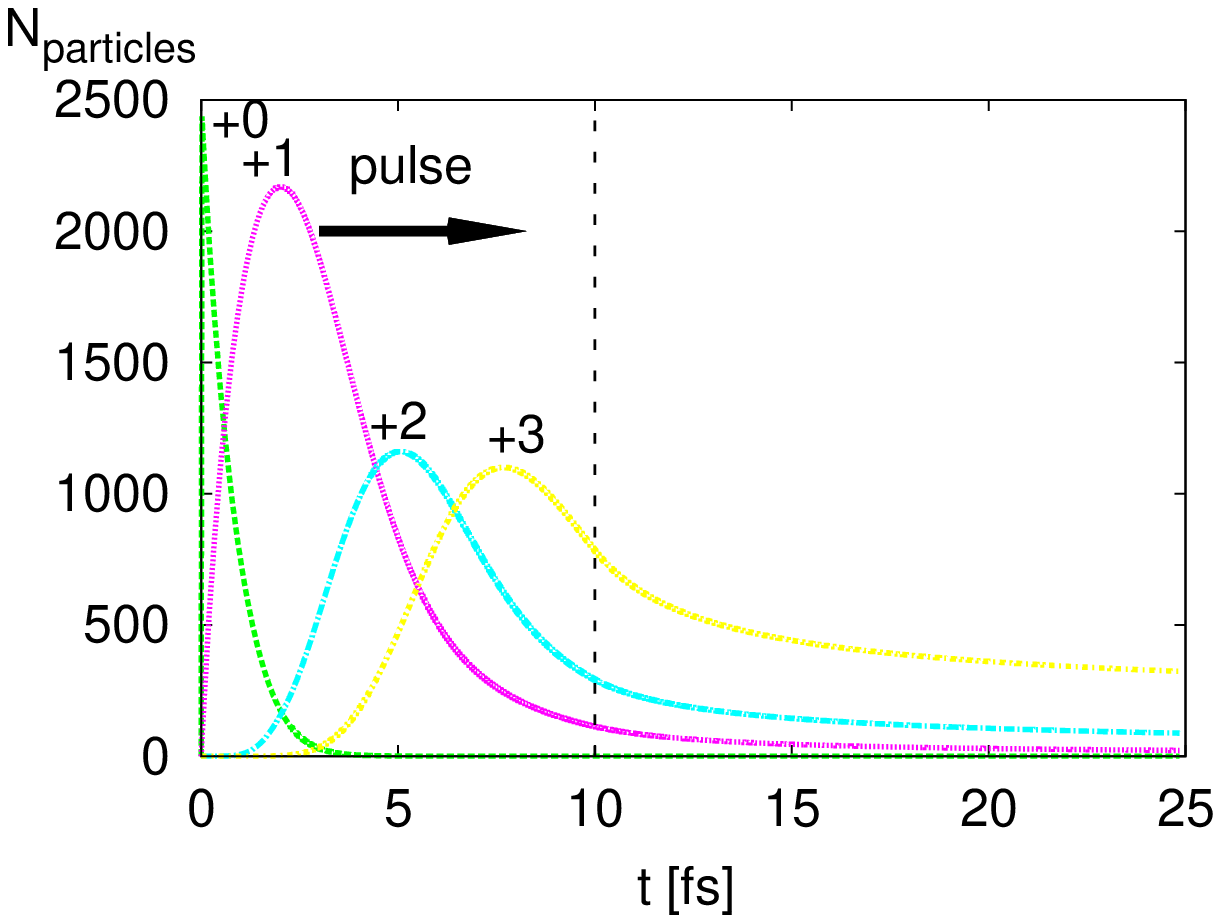}
b)\epsfig{width=7cm, file=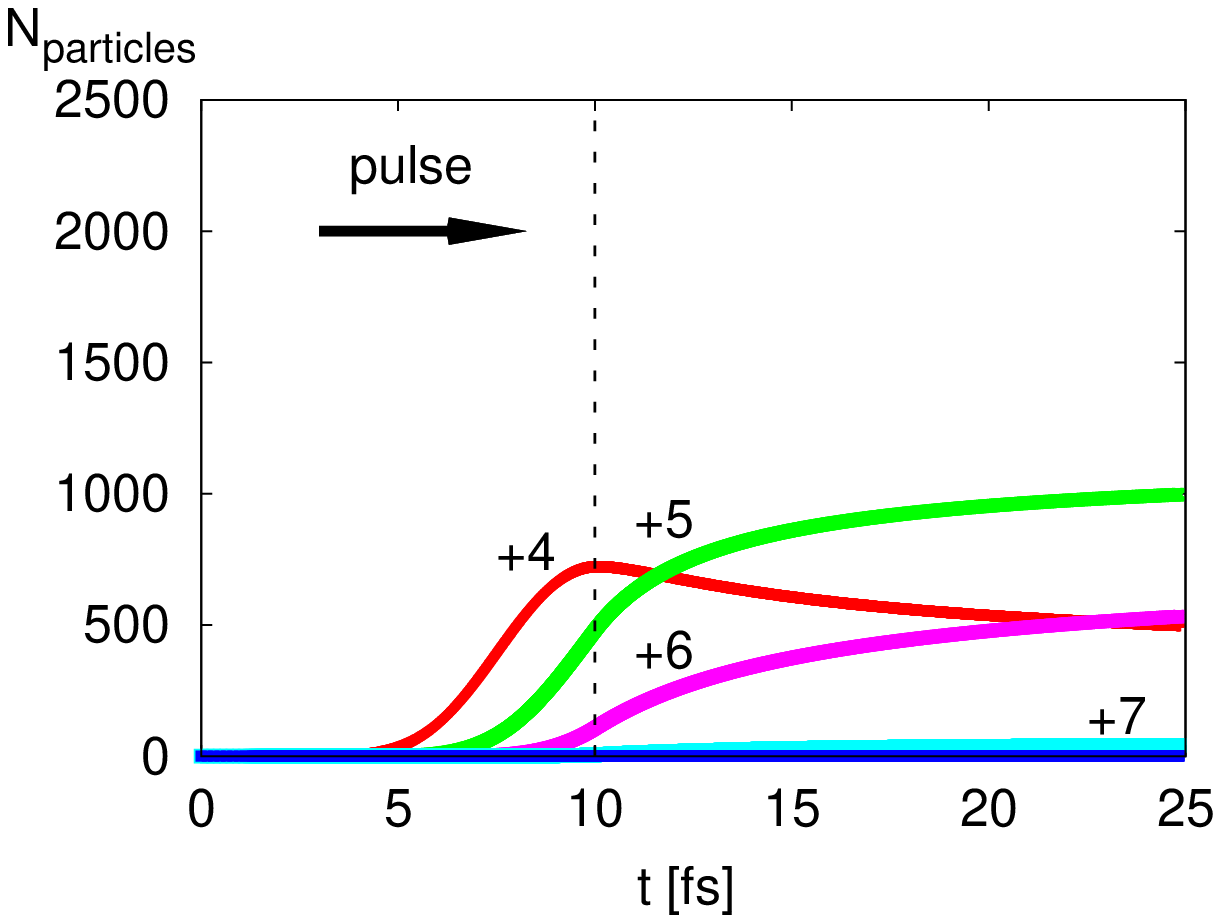} }
\caption{Creation of ions within the irradiated cluster: 
a) atoms and ions of charges, $i=1-3$, b) ions of charges, $i=4-7$. Ions of higher charges are created late in the exposure.}
\label{ions}
\end{figure}

\begin{figure}
\vspace*{0.5cm}
\centerline{\epsfig{width=7cm, file=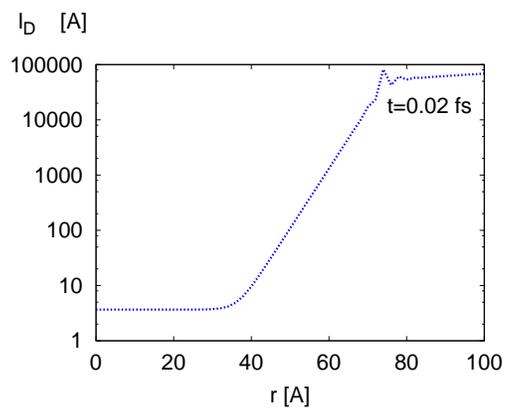}}
\caption{Debye length, $l_D$, calculated with electrons at $t=0.02$ fs within the whole simulation box. Debye length is small comparing to the cluster size of radius, $R=36$ \AA. This is one of the plasma signatures.
}
\label{deb}
\end{figure}

\noindent
{\bf Collisional ionization}\\
Heated photoelectrons can collide with ions, releasing secondary electrons: $e(E_e)+Xe^{+q} \rightarrow e(E_e^{\prime}) + e(E_{sec})+ Xe^{+(q+1)}$. These secondary electrons, $e(E_{sec})$,
will be also heated, and they can collide with other ions, releasing
more electrons. This initiates cascading processes \cite{ziaja2}. 
Due to the hierarchy of ionization cross sections, ions of higher charges are created consecutively, and the highest charges are 
created at latest in the exposure (Fig.\ \ref{ions}).
Three important factors influence the collisional ionization rate:

(i) Screening and ion environment within the plasma. Early in the ionization phase the plasma is formed. Fig.\ \ref{deb} shows the Debye length calculated with electrons at time $t=0.02$ fs of the exposure. 
At this time the cluster interior is a plasma with the Debye length, $l_d\sim 3$ \AA, much less than the cluster size, $R\sim 36$ \AA. With 
the increasing number of electrons, the Debye length decreases down to 
$l_d\sim 1$ \AA. Including the effect of electron screening and cluster environment on the ionization potentials leads to the increased production
of higher ion charges as compared to the case, where collisional cross-sections are calculated with the potentials of isolated atoms. As the shifts of the ionization potentials depend on the electron density and electron temperature, and these parameters change with time, this will affect the energy absorption during collisional ionizations, leading to non-linear effects.

(ii) Heating rate. Results of this and previous simulations \cite{ziajab1} show that the maximal ion charge created within the sample strongly depends on the heating rate applied.

\begin{figure}
\vspace*{0.5cm}
\centerline{a)\epsfig{width=6cm, file=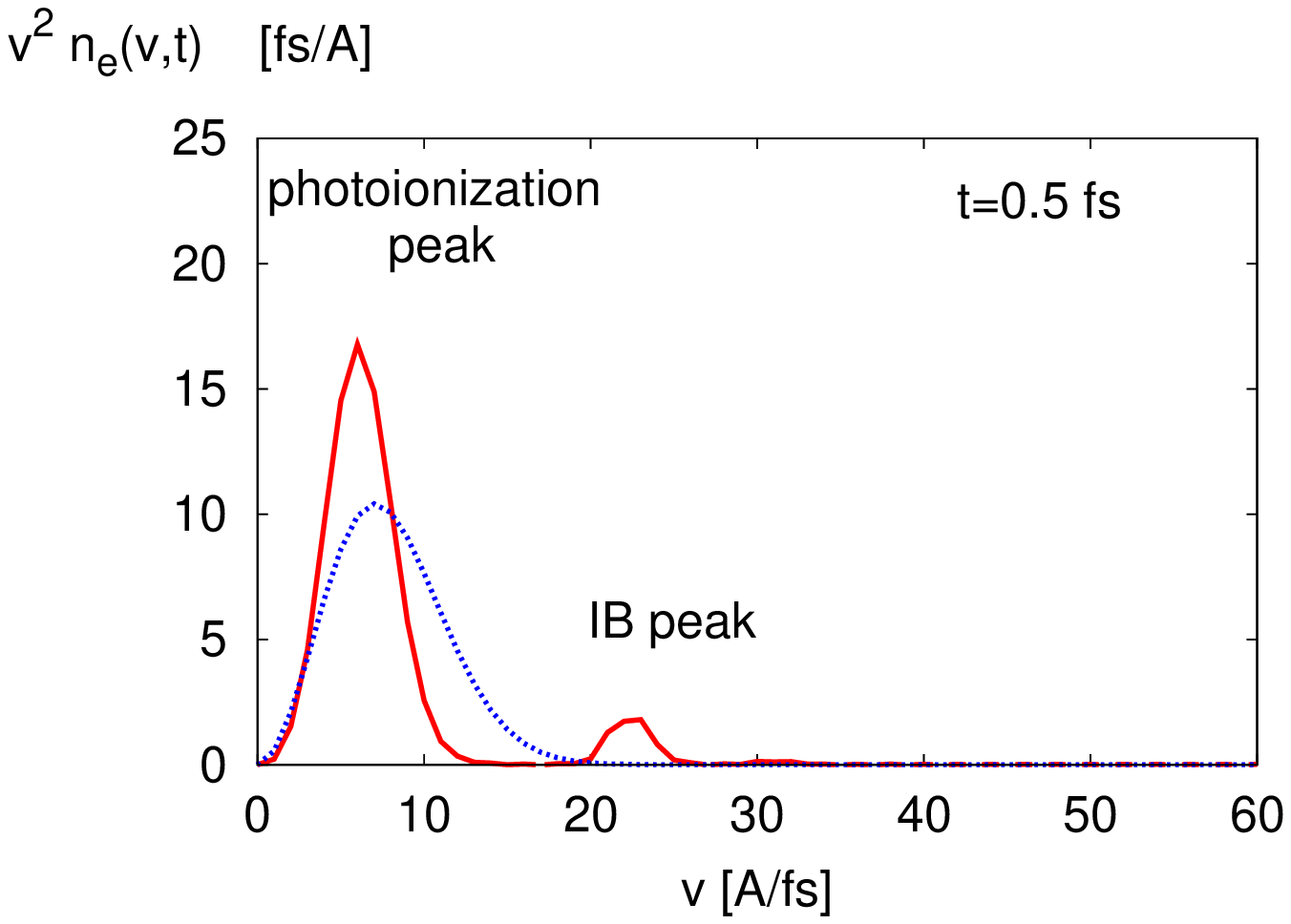}
b)\epsfig{width=6cm, file=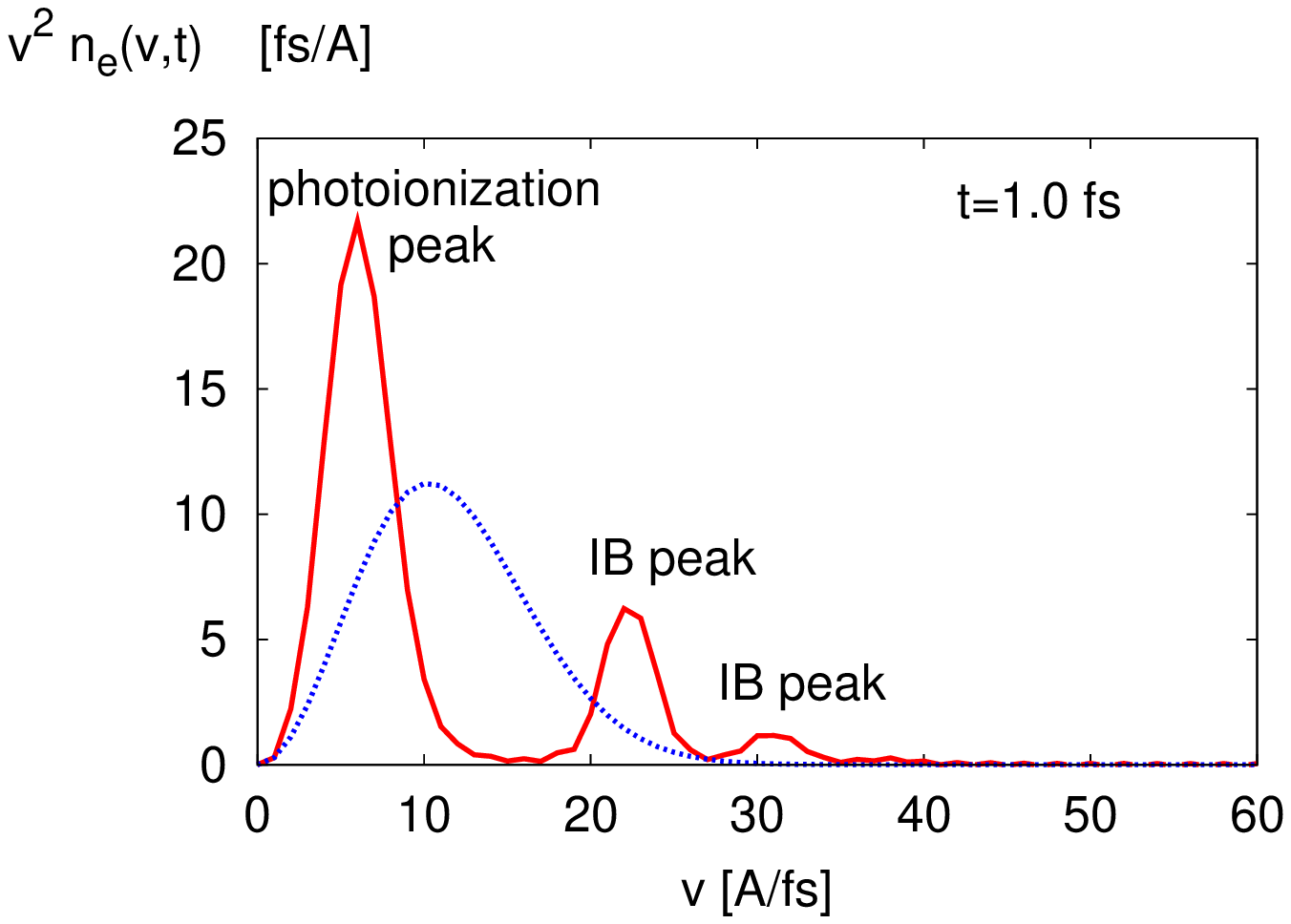} }
\vspace{0.5cm}

\centerline{c)\epsfig{width=6cm, file=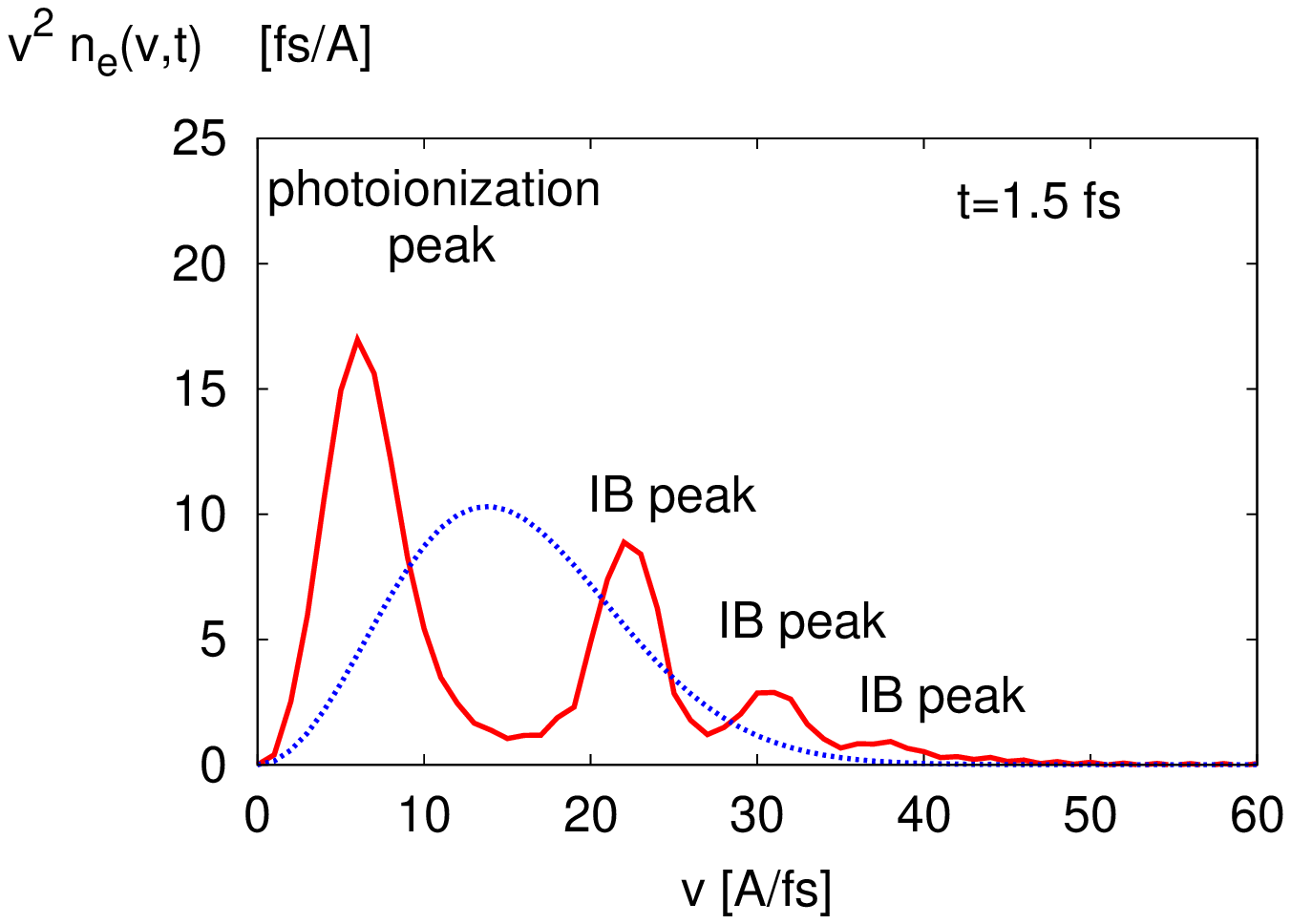}
d)\epsfig{width=6cm, file=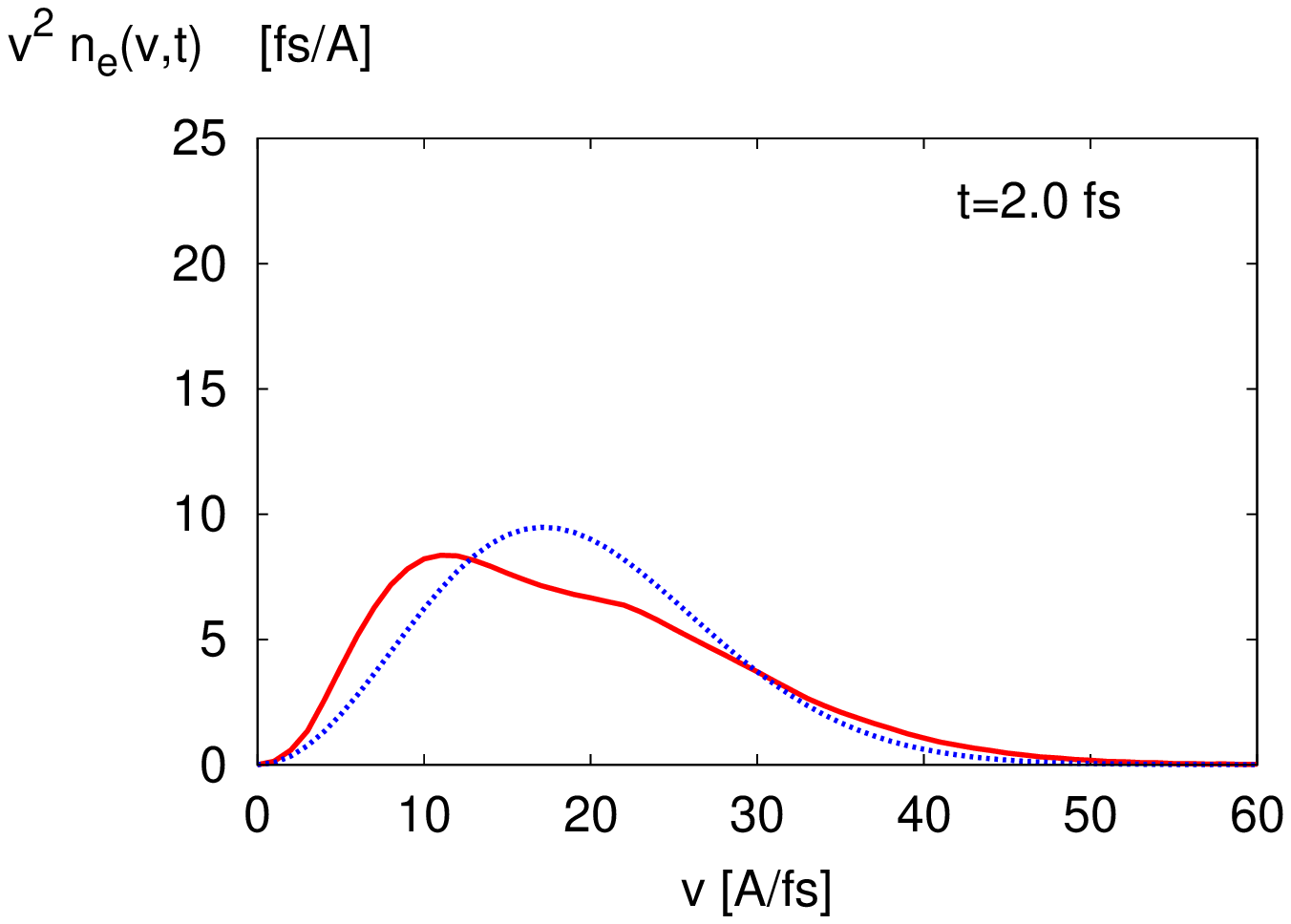} }
\vspace{0.5cm}

\centerline{e)\epsfig{width=6cm, file=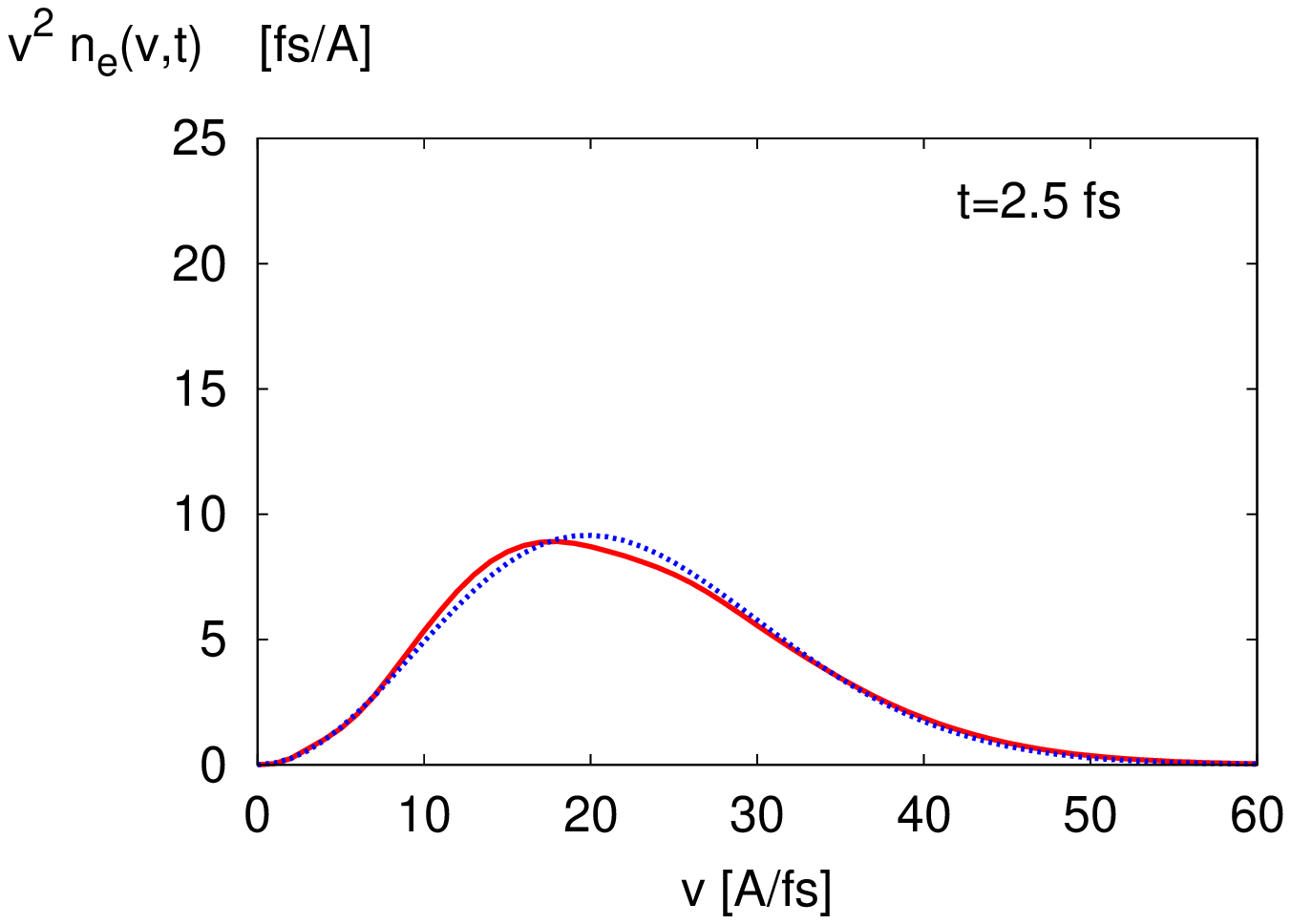}
f)\epsfig{width=6cm, file=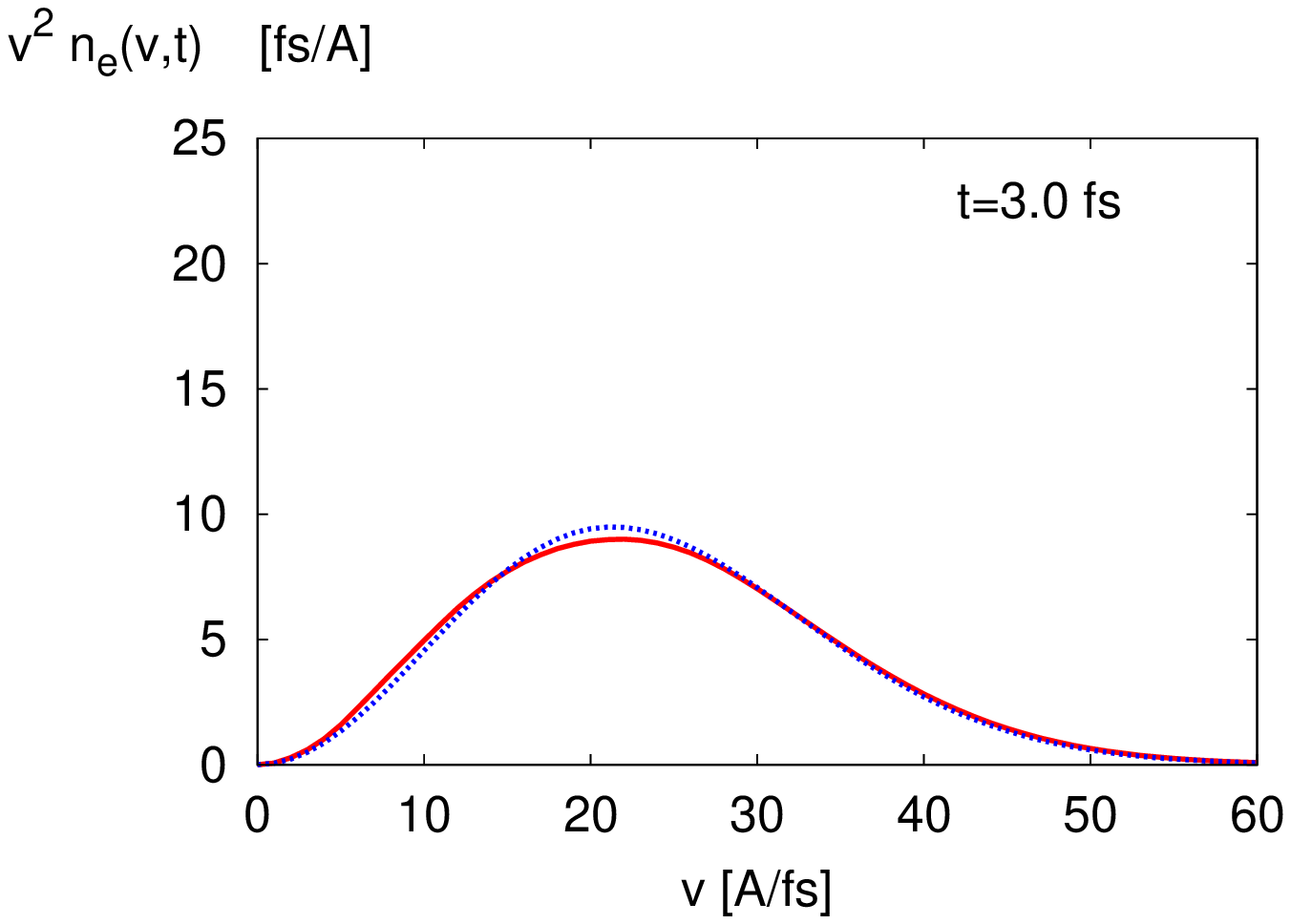} }
\vspace*{0.5cm}
\centerline{g)\epsfig{width=6cm, file=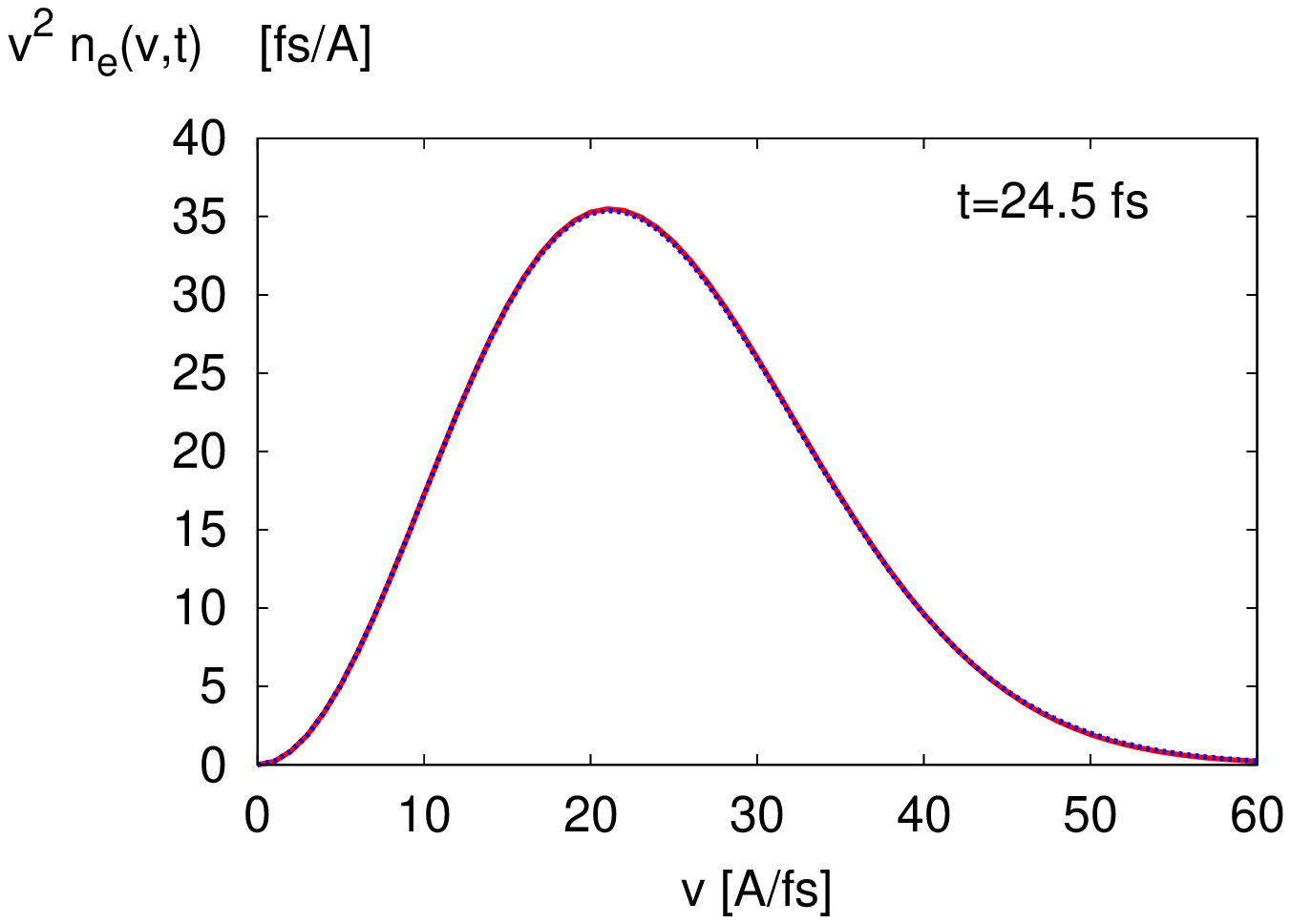}}

\caption{Effect of the IB heating rate and the shielded electron-electron interactions on the electron velocity distribution. Solid curve shows the electron distribution, dashed curve shows Maxwell-Boltzmann distribution obtained with the instantaneous temperature and the electron density estimated at a given time within the cluster. Evolution of the
electron density is shown at times, $t=0.5, 1.0, 1.5, 2.0, 2.5, 3.0$ fs of the exposure. After $3$ fs the shape of electron distribution approaches that one of the Maxwell-Boltzmann distribution. At later times (plot (g)),
after the pulse is finished and electrons are not longer heated, these 
two distributions overlap.}
\label{term}
\end{figure}

(iii) Shielded electron-electron interactions. They strongly influence the distribution of energy among plasma electrons. Fast thermalization
induced by this interaction (local thermalization timescale $\leq 3$ fs in the simulated case) cuts the tail of high electron energies (Fig.\ \ref{term}). We checked that this effect delays the appearance of higher ion charges within the sample, comparing to the case, where the shielded electron-electron interactions are not treated (not shown).

\begin{figure}
\vspace*{0.5cm}
\centerline{\epsfig{width=4cm, file=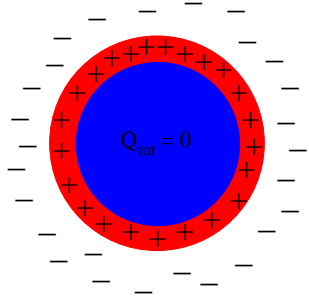}
}
\caption{Schematic plot of charge distribution within an irradiated
large cluster at the end of the ionization phase. 
Positively charged outer shell coats the neutral cluster core of a net charge equal to zero. Thermalized electrons slowly escape from the cluster.}
\label{large}
\end{figure}

\begin{figure}
\vspace*{0.5cm}
\centerline{a) \epsfig{width=7cm, file=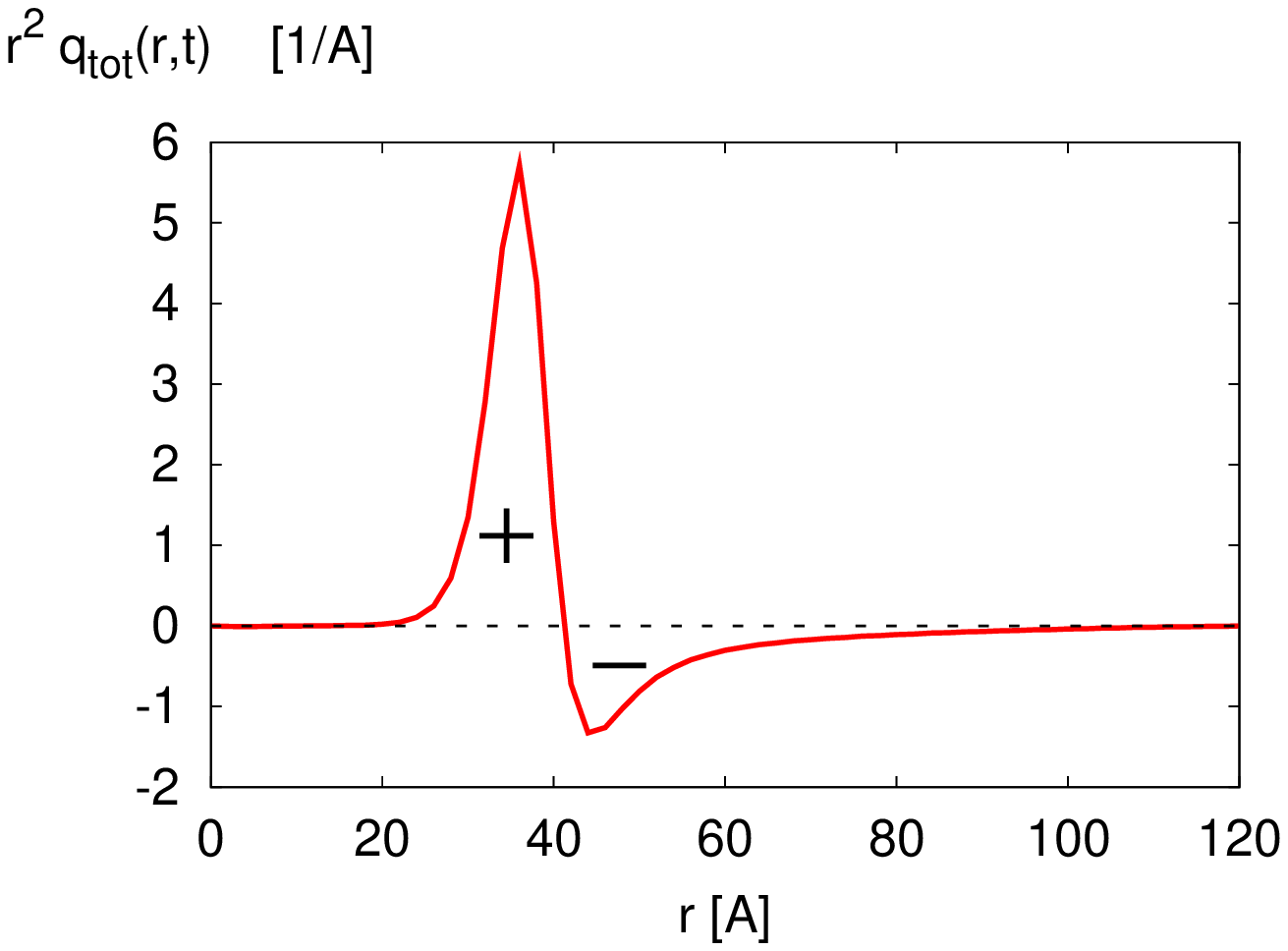}
b) \epsfig{width=7cm, file=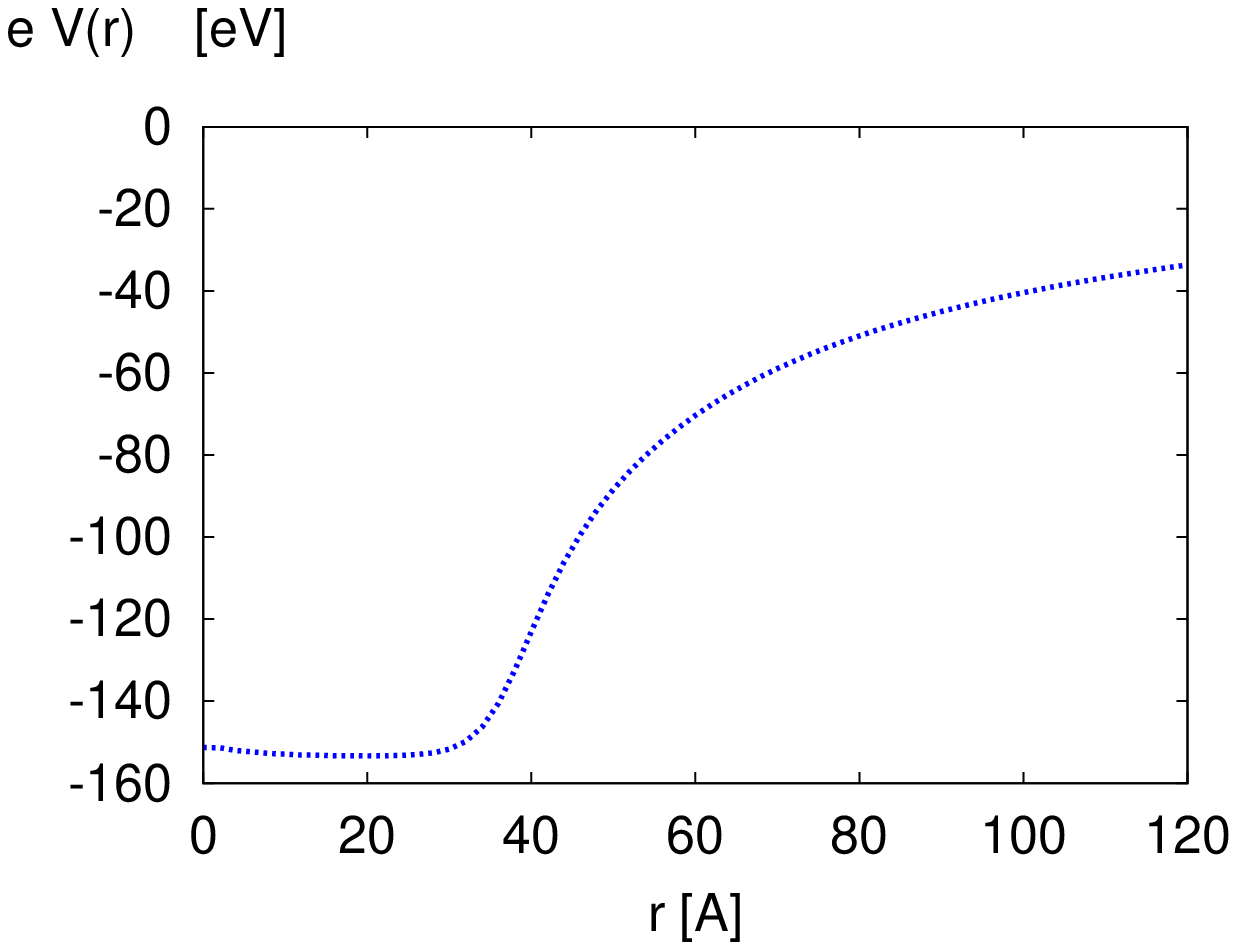}}
\caption{Electrons and ions during the ionization phase: 
a) formation of the outer shell of ions. The charge density is defined as $q_{tot}(r,t)=\sum_{i=1}^{N_J}\, i\cdot n_i(r,t) -n_e(r,t)$, and b) electrostatic attractive potential felt by electrons.}
\label{pot}
\end{figure}

\begin{figure}
\vspace*{0.5cm}
\centerline{\epsfig{width=7cm, file=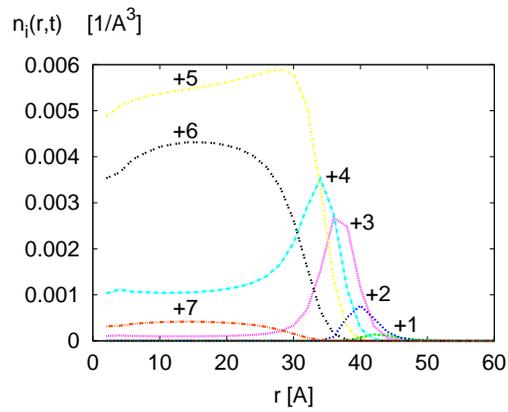}}
\caption{Inhomogeneous spatial distribution of ion charge: atom and ion densities, $n_i(r,t)$ ($i=0-8$), are plotted as functions of the distance from the centre of the cluster. These densities were recorded at the end of the ionization phase. Interior of the cluster is dominated by highest ion charges. Low charges can be found only at the edge of the cluster.}
\label{out}
\end{figure}

\newpage
\noindent
{\bf Charge distribution}\\
At the end of the pulse the charge distribution within the sample
evolves into a characteristic layer structure consisting of a neutral core and of positively charged outer shell (Figs.\ \ref{large}, \ref{pot}a). The interior of the cluster (core) is dominated by ions of the highest charges (Fig.\ \ref{out}), however the net charge of the core remains equal to 0. This is due to the quasi-free electrons bound within the core. The positively charged surface layer consists of ions of various charges. This inhomogeneous spatial distribution of charges is created in the following way. During the irradiation the most energetic electrons escape from the sample. The remaining ions create a Coulomb potential that keeps the slower electrons within the sample (Fig.\ \ref{pot}b). These electrons move freely within the cluster, and have the largest velocities when they are far from the cluster edge. At the edges electrons are stopped by the ion potential. As a result they do not ionize efficiently at the cluster edge but they do ionize the interior of the cluster. Therefore the highest ion charge is created within the core.

However, let us stress the point, that the ion distributions observed within the cluster at the end of the ionization phase will not correspond to those recorded by the detector during experiments. 
Ions from the outer shell will be the first ones to escape from the sample, and they will reach the time-of-flight (TOF) detector with unchanged charge distribution. In contrast, the cluster core at the end of ionization phase is a dense system of quasi-free electrons and ions. Recombinations and ionizations (to and from excited states) still occur within the sample. During the long picosecond {\bf expansion phase} the charges within the core will have enough time for the efficient recombination. As a result, the remnants of the core will be weakly charged or neutral. They will reach the detector late, during the hydrodynamic expansion of the core. The charge distribution recorded in the TOF spectrum will then be modified by increasing the participation of lower charges. The mechanism proposed here should be quantitatively verified with an expansion code, e.g. a hydrodynamic code. This is, however, beyond the scope of the present study.

\noindent
{\bf Global parameters}\\
Finally we discuss global parameters obtained with our model 
(Figs.\ \ref{global}a-c). Ionizations (from ground states) saturate within $\sim 10$ fs after the pulse was switched off. The electron temperature increases during the pulse. This is due to the heating of electrons within the cluster. The temperature decreases rapidly after the pulse is switched off, as the system cools fast during the collisional ionizations. After the saturation of ionizations the electron temperature decreases with time much slower. This effect is due to the slow escape of the thermalized electrons from the cluster. 

The total energy of the system increases non-linearly with time during the pulse. Photoionization and inverse bremsstrahlung are two mechanisms of the energy absorption. After the pulse, the total energy is conserved at the considered short simulation timescales.

Our simulation has been stopped at the end of ionization phase, i.e. after the saturation of ionization was observed. Although the system has not undergone the full evolution yet, we can derive some physical predictions from the simulation results. They are: (i) maximal and average ion charge observed, (ii) distribution of ions within the outer shell, (iii) limits for the total absorbed energy per atom, (iv) thermalization timescales. Predictions (i)-(iii) can be compared to the experimental data.


\section{Comparison to experimental data}

For further analysis we estimate the total amount of pulse energy transferred through a unit surface during the pulse. This is the time integrated energy flux, $F$. For a flat pulse of intensity, $I$, and duration, $\Delta t$, it takes a simple form: $F=I\cdot \Delta t$.

We compare the results of our simulations to the experimental data 
from the first experiment performed with FLASH at DESY in 2001. Ion fractions and average energy absorption estimated with averaged TOF spectra were recorded in this experiment at five different pulse energy fluxes: $F=0.05, 0.3, 0.84,
1.25,1.5$ J/cm$^2$ \cite{desy2}. The error in estimation of the value of $F$ could be up to a factor 5. We remind here that the TOF detector could record charged particles (ions) only. There are no data on the neutrals available from these measurements. Experimental predictions that we use here were obtained with the averaged integrated intensities recorded at TOF detectors. For fluxes, $F=0.84, 1.25$ J/cm$^2$, those intensities were weighted with relative geometric acceptances and the MCP detector efficiencies. For fluxes, $F=0.05, 0.3, 1.5$ J/cm$^2$ only unweighted data are available.

We simulated the non-equilibrium phase of the evolution of $Xe_{2500}$ clusters exposed to single flat VUV pulses of a fixed flux, $F$, but of various intensities and pulse durations. Intensities and pulse duration were chosen in order to match the condition: $I \cdot \Delta t=F$. Pulse intensity was $\leq 10^{14}$ W/cm$^2$ and pulse length, $\Delta t \leq 50$ fs.  
The predictions obtained from different events were then averaged over the number of events. This procedure enabled us to account for the non-linear response of the system to the various pulse lengths and pulse intensities at higher radiation fluxes. This scheme followed the experimental data analysis: experimental data were obtained after averaging the single shot data obtained with various FEL pulses of a fixed radiation flux. 

First, we show the simulation results obtained with standard IB rates \cite{krain1,krain}. These rates estimated the heating of the quasi-free electrons during their scattering on the Coulomb potentials of point-like ions. They were calculated separately for slow and fast electrons. We also included the modification of the ionization potentials by plasma environment. The hybrid model \cite{murillo} and an estimate of the effect of the surrounding ions (discussed in detail in the preceding section) were used for calculating the energy level shifts within the plasma.
  
\begin{figure}
\vspace*{0.5cm}
\centerline{a)\epsfig{width=6cm, file=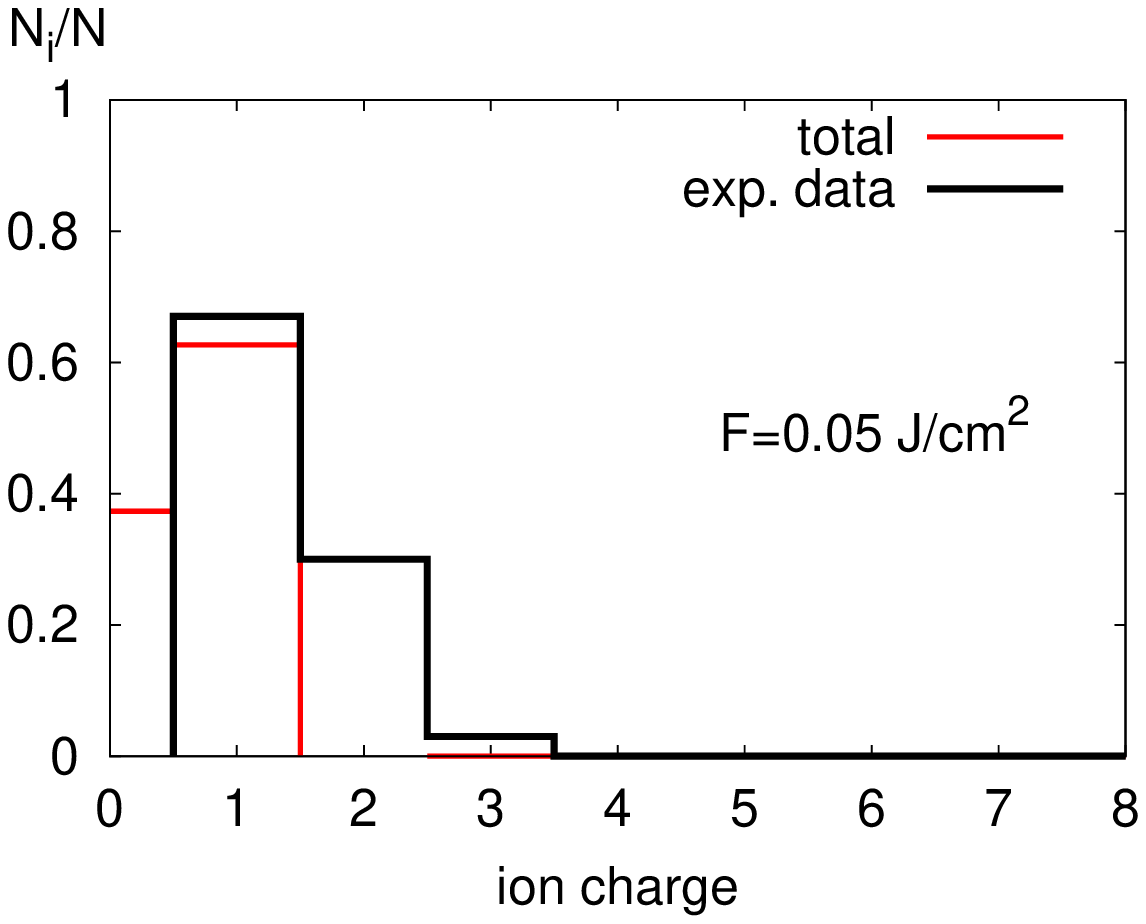}
b)\epsfig{width=6cm, file=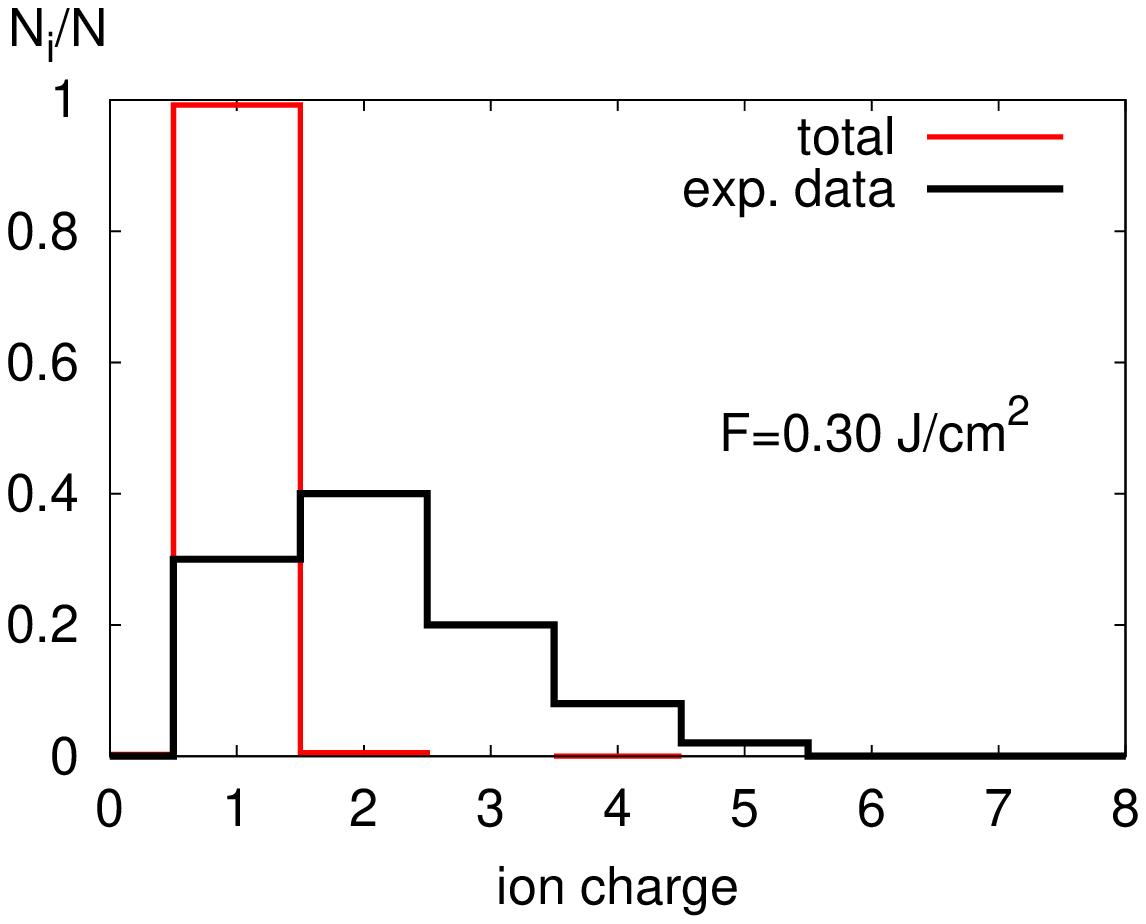} }
\vspace{0.5cm}

\centerline{c)\epsfig{width=6cm, file=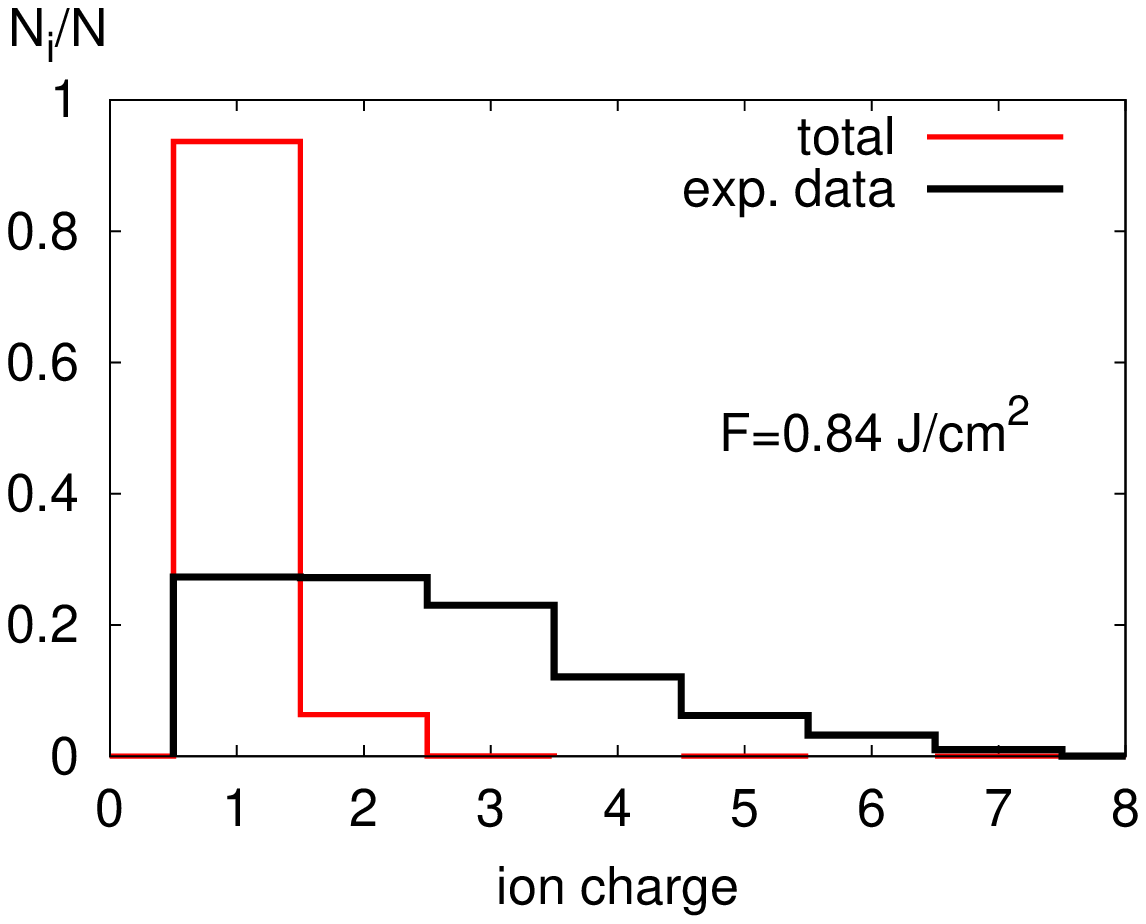}
d)\epsfig{width=6cm, file=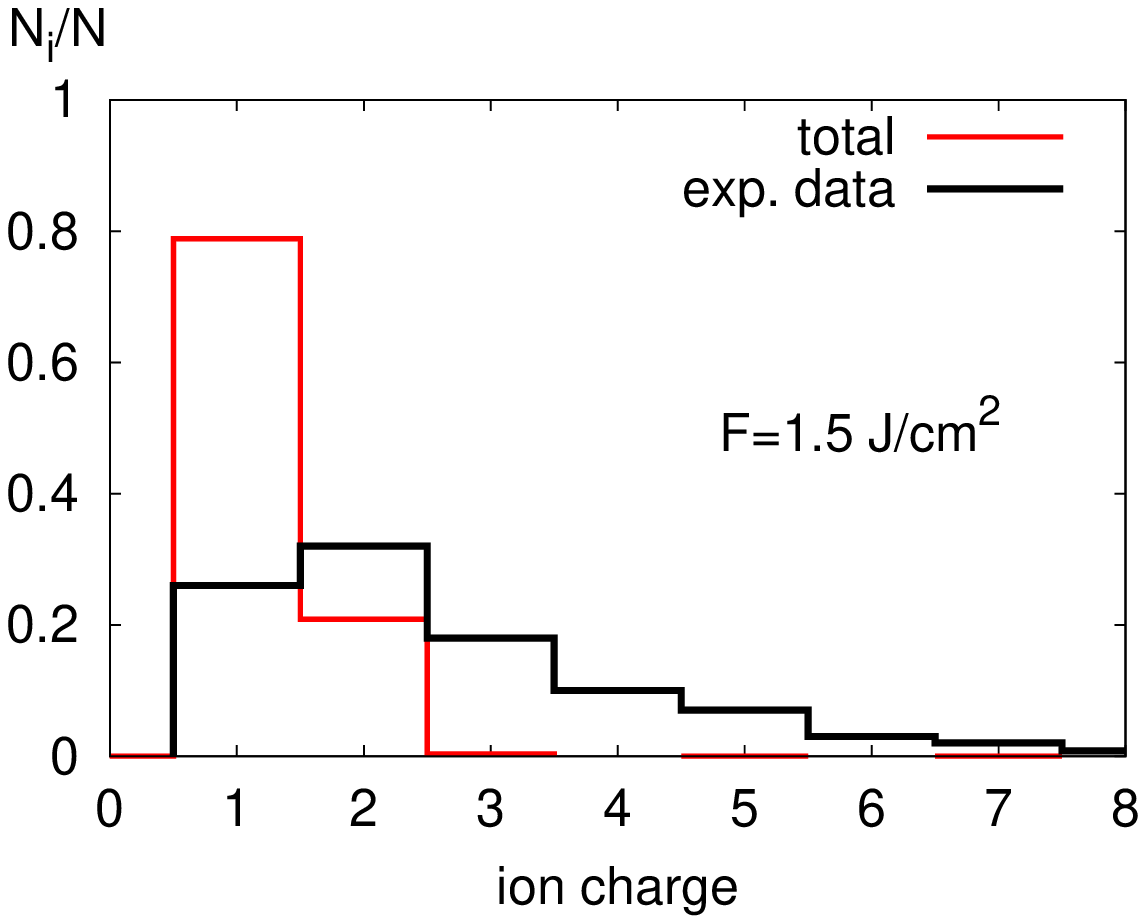} }
\vspace{0.5cm}

\caption{{\it Results with the standard IB rate and the plasma modified atomic potentials:} Ion fractions within the irradiated xenon cluster at the end of ionization phase calculated within the whole cluster. They are compared to the experimental data. In each case a)-d) these clusters were irradiated with pulses of different intensities 
($\leq 10^{14}$ W/cm$^2$) and lengths ($\leq 50$ fs) but of a fixed flux. The results obtained were then averaged over the number of pulses. Irradiation at four different radiation fluxes: a) $F=0.05$ J/cm$^2$, b) $F=0.3$ J/cm$^2$, c) $F=0.84$ J/cm$^2$, and d) $F=1.5$ J/cm$^2$ was considered.}
\label{chargek}
\end{figure}

Below we show: (i) ion fractions obtained with the experimental data, (ii) ion fractions obtained within the whole cluster with our model (Fig.\ \ref{chargek}). At the fluxes, $F=0.05, 0.3$ J/cm$^2$, only single charged
ions were observed. At higher fluxes $F=0.84, 1.5$ J/cm$^2$ Xe ions up to 
$+2$ could be detected. These predictions are in disagreement with the experimental findings that predict much higher charge states at higher radiation fluxes. Obviously, these heating rates were too low to lead to the creation of higher charges within the sample, at least with the modification of ionization potentials obtained with the hybrid screening model and the barrier lowering
modelled as described in the preceding section.

\begin{figure}
\vspace*{0.5cm}
\centerline{a)\epsfig{width=6cm, file=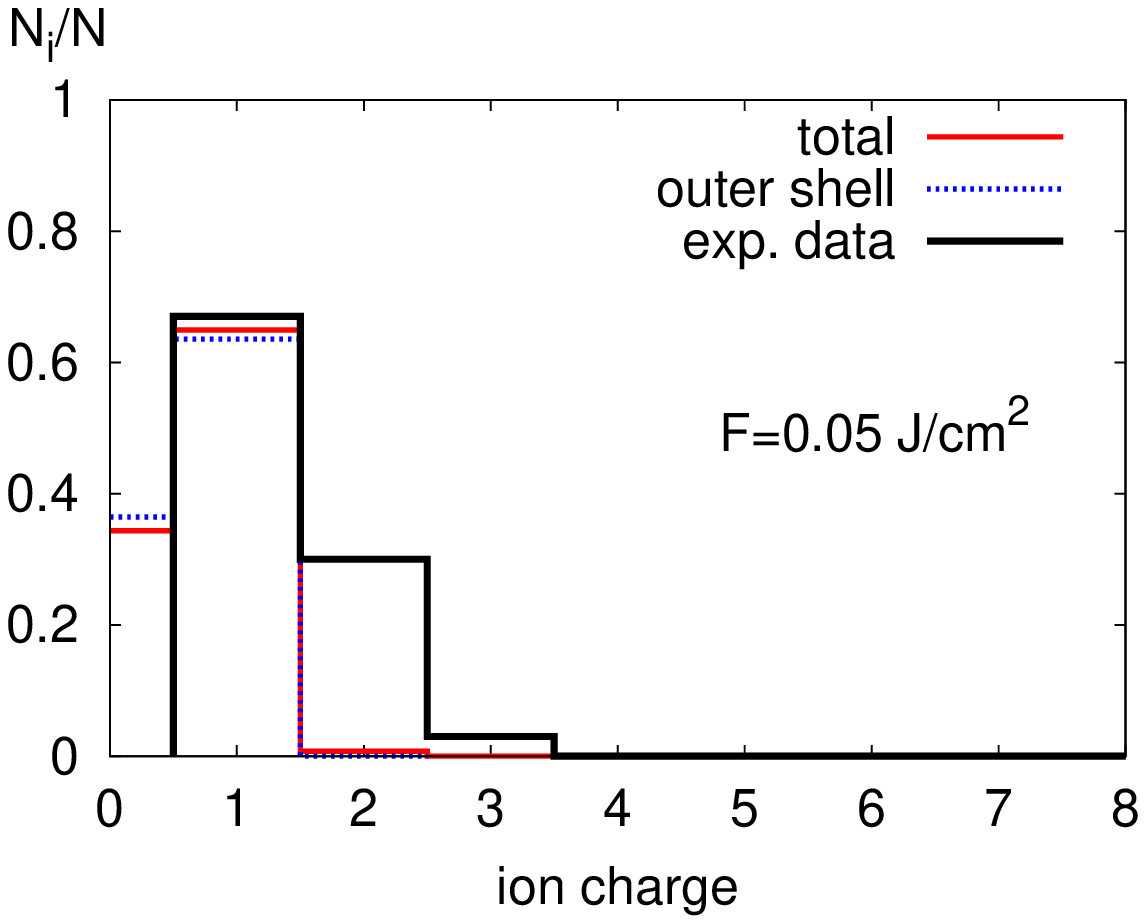}
b)\epsfig{width=6cm, file=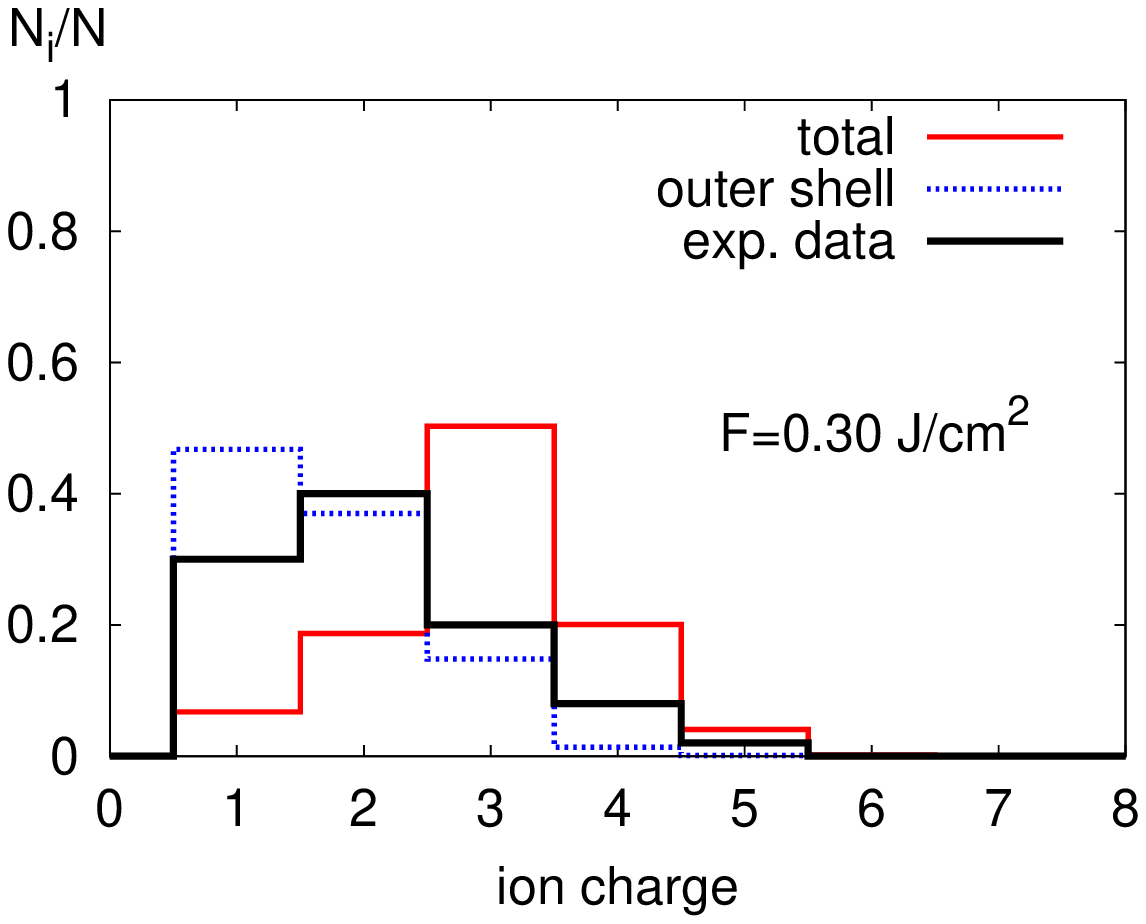} }
\vspace{0.5cm}

\centerline{c)\epsfig{width=6cm, file=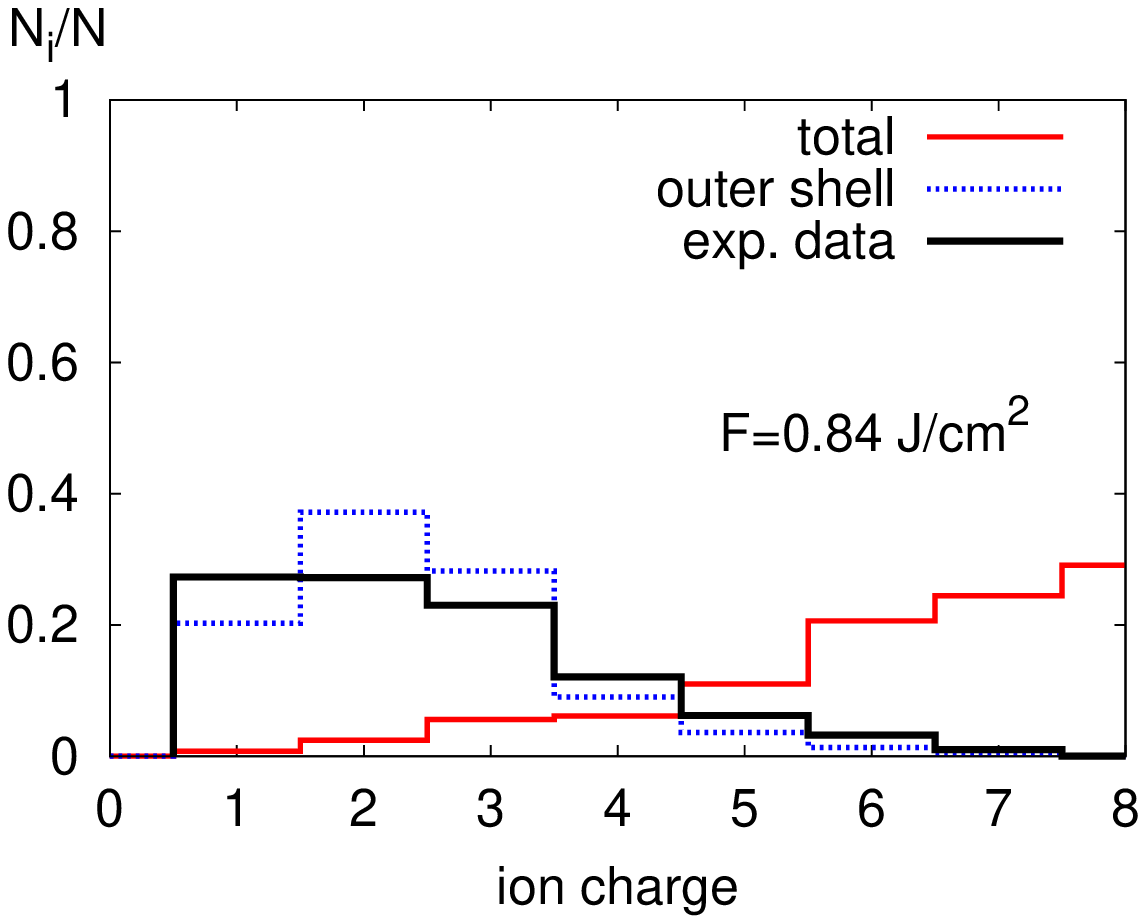}
d)\epsfig{width=6cm, file=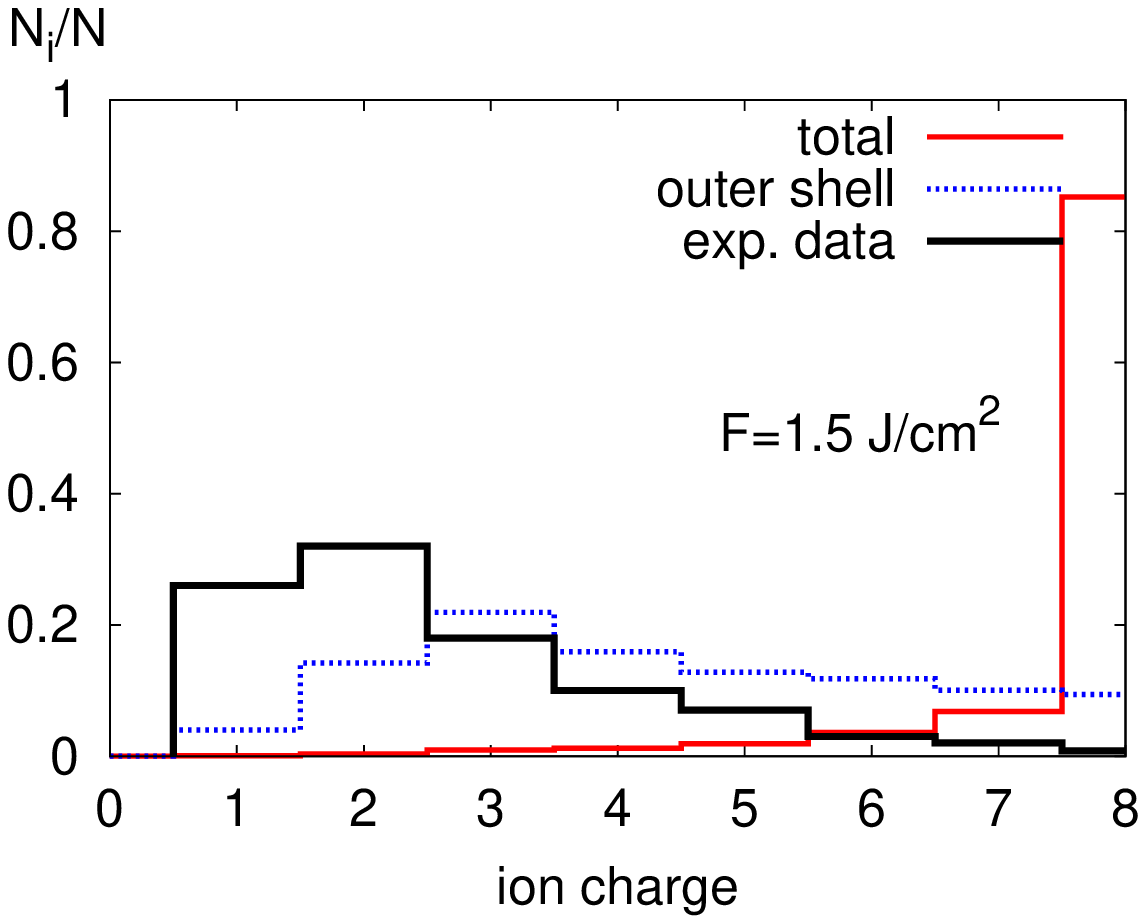} }
\vspace{0.5cm}

\caption{{\it Results with the enhanced IB rate and the plasma modified atomic potentials:} Ion fractions within the irradiated xenon cluster at the end of ionization phase calculated within the whole cluster and within the outer shell. They are compared to the experimental data. In each case a)-d) these clusters were irradiated with pulses of different intensities 
($\leq 10^{14}$ W/cm$^2$) and lengths ($\leq 50$ fs) but of a fixed flux. The results obtained were then averaged over the number of pulses. Irradiation at four different radiation fluxes: a) $F=0.05$ J/cm$^2$, b) $F=0.3$ J/cm$^2$, c) $F=0.84$ J/cm$^2$, and d) $F=1.5$ J/cm$^2$ was considered.}
\label{charge}
\end{figure}

Second, we show the results of the simulations performed with the enhanced IB rate as proposed in Ref.\ \cite{santra} and with the plasma shifted atomic energy levels. These rates were estimated with the effective atomic potential. Below we show: (i) the plots of the ion fractions obtained with the experimental data, (ii) ion fractions obtained within the whole cluster with our model, (iii) ion fractions obtained within the surface layer (outer shell) with our model (Fig.\ \ref{charge}). 

At the lowest flux, $F=0.05$ J/cm$^2$, we obtain a large discrepancy with the data. In the experiment ions of charge up to +3 were found. In the simulation we find ions up to +2 at most. Also, ion fractions are very different, e.g. the high participation of neutrals predicted within our model cannot be verified with experimental data. Experimental data on the charge distribution at $F=0.05$ J/cm$^2$ can be well fitted with our model at $F=0.11-0.13$ J/cm$^2$ (not shown). This is still within the experimental error of the estimation of the radiation flux.

At the flux, $F=0.3$ J/cm$^2$, experimental ion fractions lay between the theoretical fraction histogram obtained within the whole cluster and that one obtained within the outer cluster shell. Maximal ion charge is found to be +5 with both experimental data and simulation results.

At higher fluxes, $F=0.84$ and $1.5$ J/cm$^2$, the ion fractions predicted within the whole cluster overestimate the experimental data. However, the distribution of ions within the surface layer follows the tendency of data, with the maximum at charge +3. 
If recombination within the cluster core would be efficient during the expansion phase, ion charge within the core should be significantly reduced. The recorded ion spectra from outer shell would then be corrected by the contributions of the weakly charged remnants of the expanded core. The total charge distributions obtained should then be in agreement with the experimental ones.

First we list our detailed predictions on ion charges. The model predictions on maximal ion charges, $Z_{max}$, follow the experimental data for higher fluxes, $F=0.3, 0.84, 1.5$ J/cm$^2$: $Z_{max}=+5$ for $F=0.3$ J/cm$^2$, $Z_{max}\geq +7$ for $F=0.84$ J/cm$^2$, and $Z_{max}= +8$ for $F=1.25, 1.5$ J/cm$^2$. For comparison, if the pulse length would be fixed to, $\Delta t=50$ fs, the radiation fluxes of $F=0.3,0.84,1.25,1.5$ J/cm$^2$ could be achieved with the following intensities, $I\sim 0.6,1.7,2.5,3\,\cdot$  $10^{13}$ W/cm$^2$.  

\begin{figure}
\vspace*{0.5cm}
\centerline{\epsfig{width=7cm, file=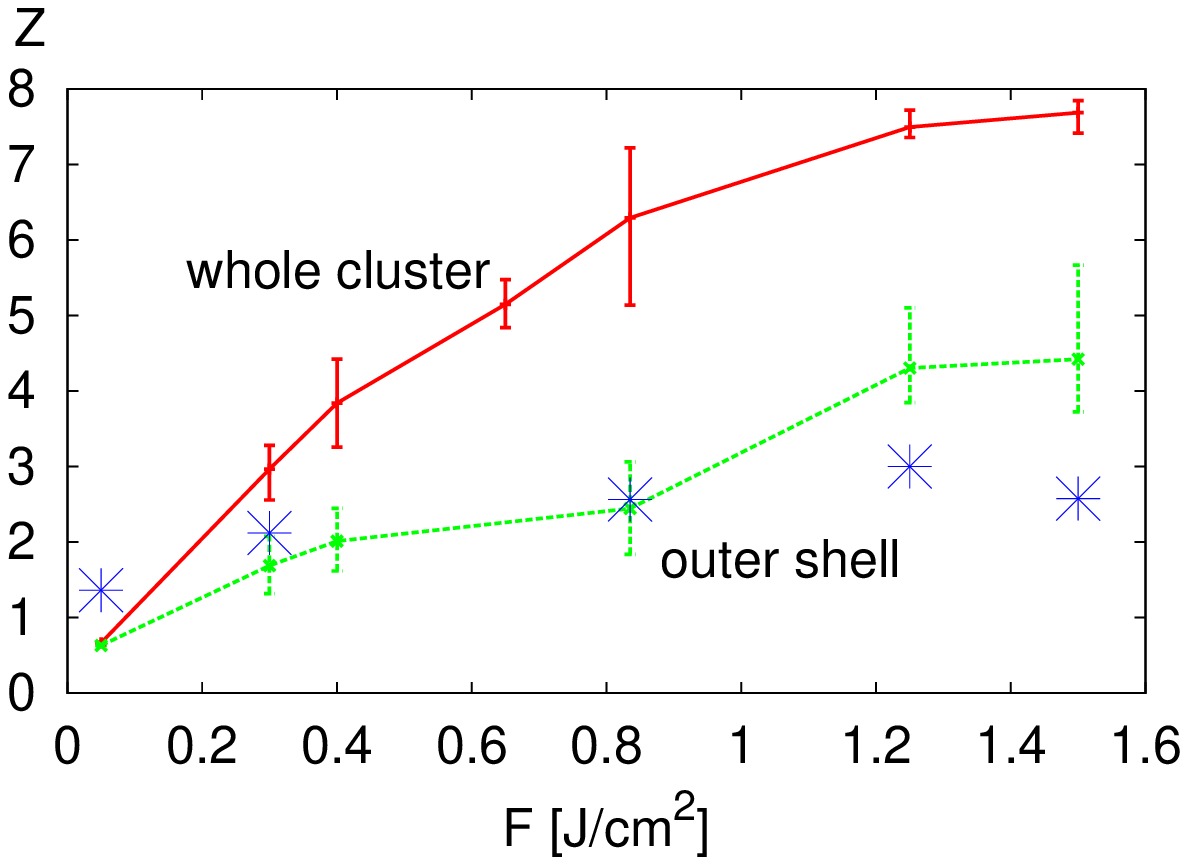}}
\caption{{\it Results with the enhanced IB rate and the plasma modified atomic potentials:} Average charge, $Z$, created within the whole irradiated cluster
(red errorbars) and within the outer shell (green errorbars) as a function of the time-integrated radiation flux, $F$. These estimates were obtained with pulses of different intensities and lengths but of the fixed flux, and then averaged over the number of pulses. Errorbars denote maximal errors. Experimental data are plotted with stars.}
\label{lad}
\end{figure}

The average charge is plotted as a function of radiation flux in Fig.\ \ref{lad}. The charges calculated within the outer shell are close to the corresponding experimental values. This indicates that recombination should be efficient during the expansion phase so that the remnants of the core are  weakly charged (or neutral). 

\begin{figure}
\vspace*{0.5cm}
\centerline{\epsfig{width=7cm, file=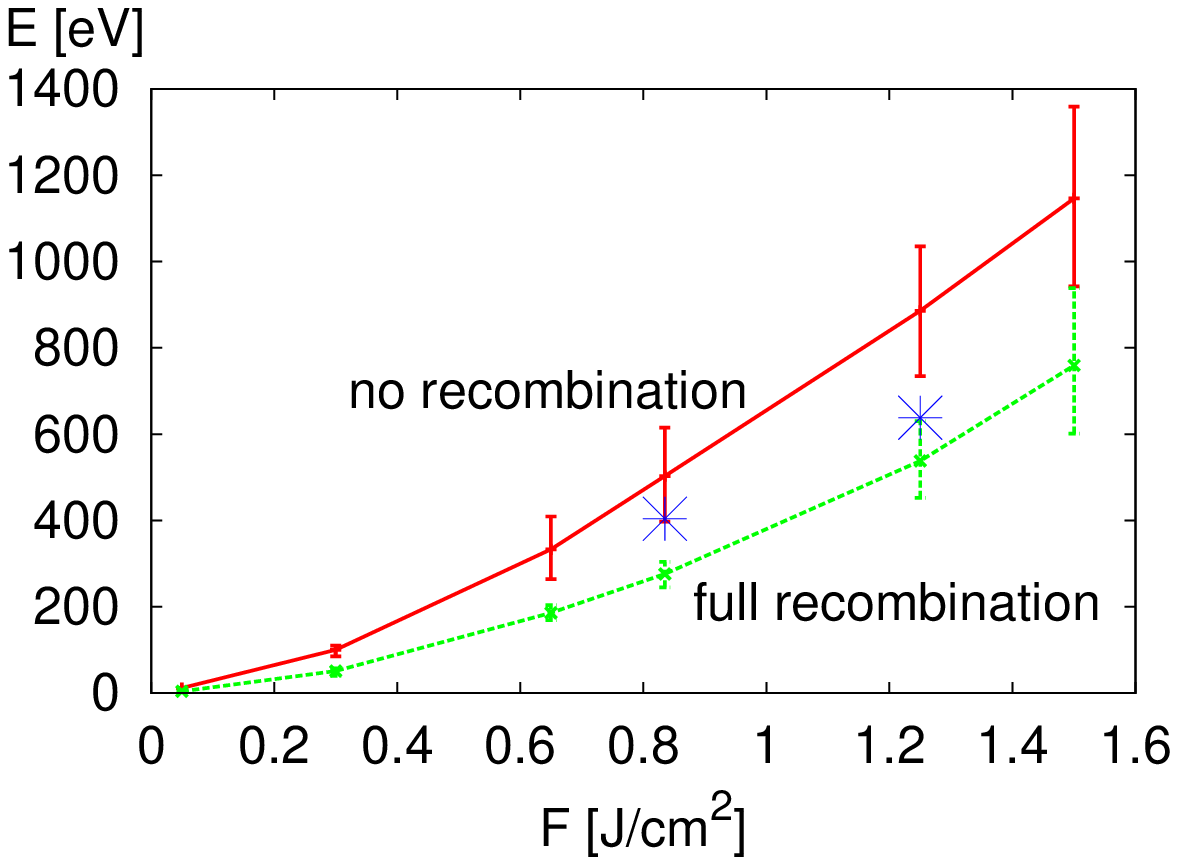}}
\caption{{\it Results with the enhanced IB rate and the plasma modified atomic potentials:} Average energy absorbed per atom, $E$, within the irradiated cluster as a function of the time-integrated radiation flux, $F$. Upper
(red errorbars)  and lower (green errorbars) limits for the absorbed energies are estimated within our model. These estimates were obtained with pulses of different intensities and lengths but of the fixed flux, and then averaged over the number of pulses. Errorbars denote maximal errors. Experimental data are plotted with stars.}
\label{energy}
\end{figure}

Below we show also the average energy absorbed per atom (estimated with our model) as a function of the radiation flux (Fig.\ \ref{energy}). 
With our model we can only obtain the upper and the lower limit for this absorbed energy. Upper limit assumes that during the further expansion of the sample no recombination processes are occurring. Lower limit gives the energy absorption estimate in case of the full neutralization of the sample during the expansion (full recombination). These limits are compared to the available experimental data on the average ejection energy per atom. The experimental data lay within the model estimates. As expected, the energy absorption shows the non-linear increase with the increasing radiation flux. 

\begin{figure}
\vspace*{0.5cm}
\centerline{a)\epsfig{width=6cm, file=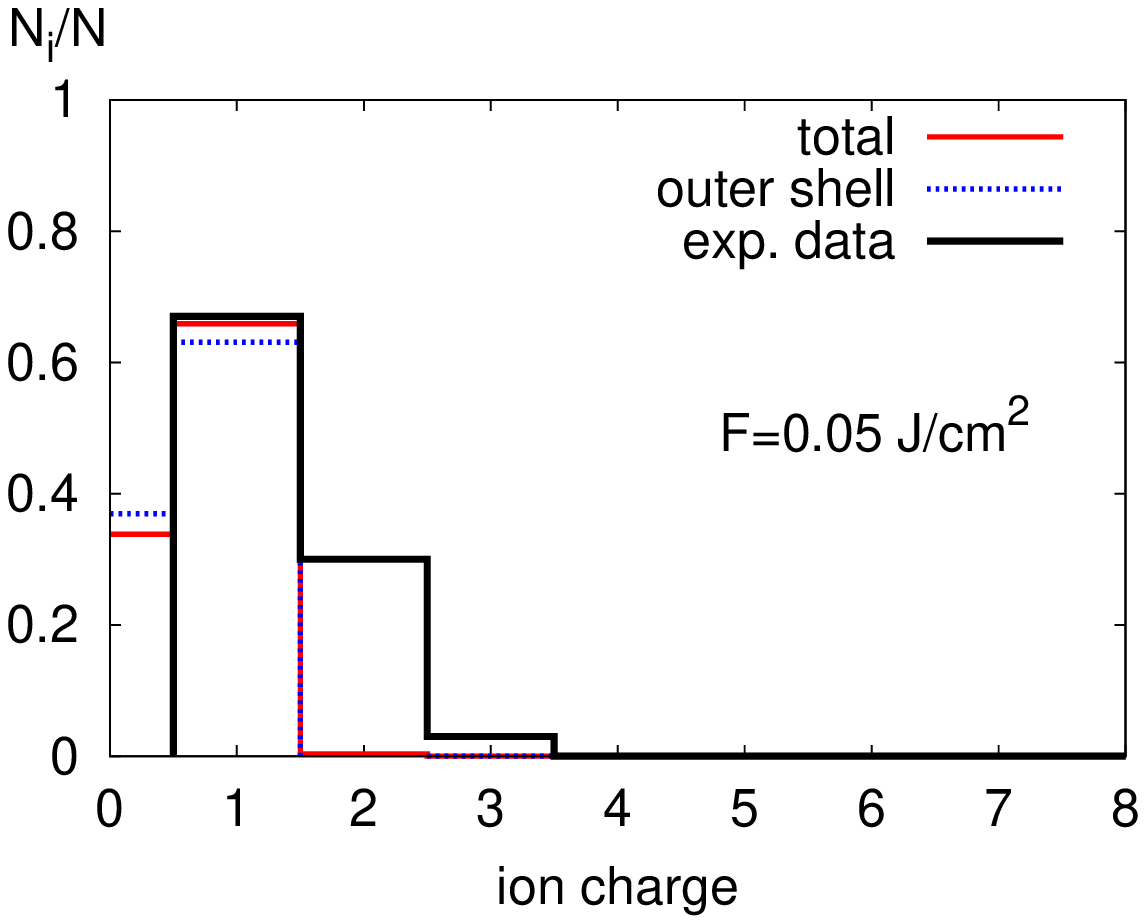}
b)\epsfig{width=6cm, file=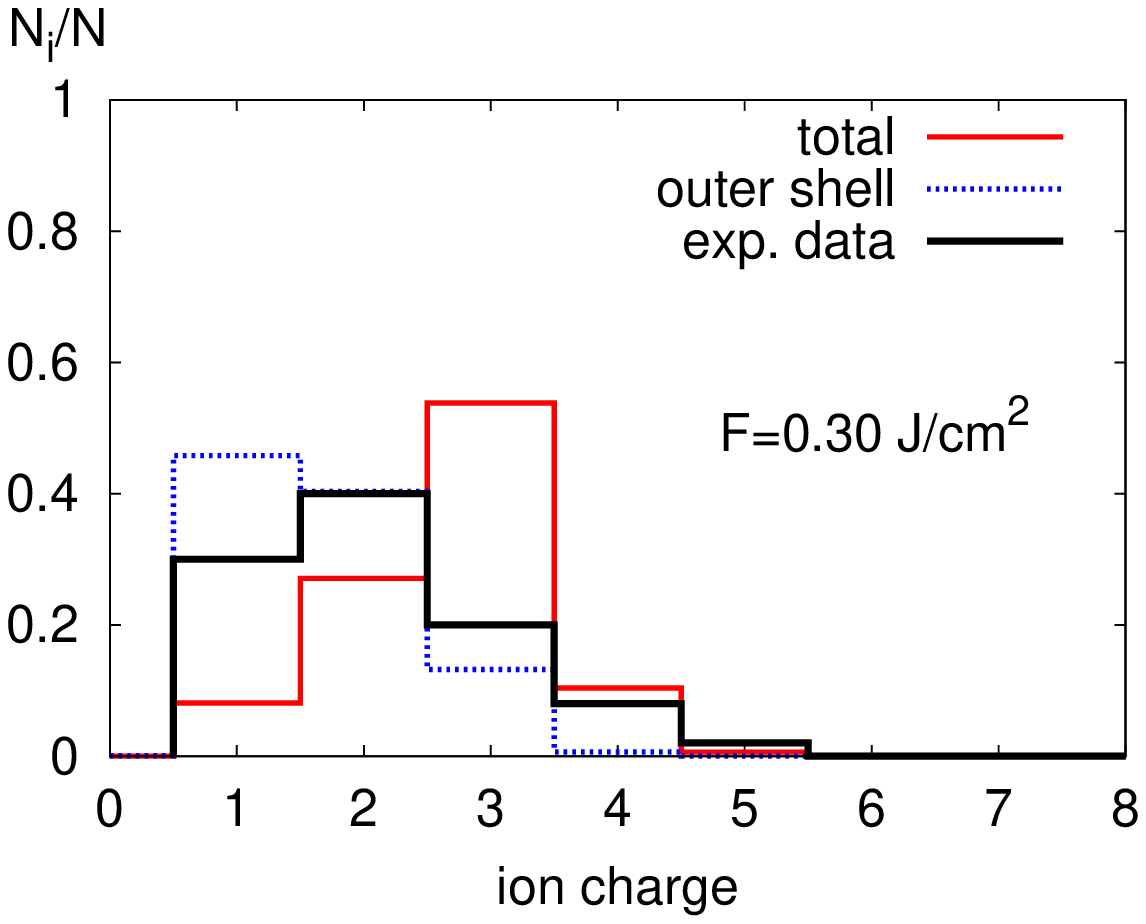} }
\vspace{0.5cm}

\centerline{c)\epsfig{width=6cm, file=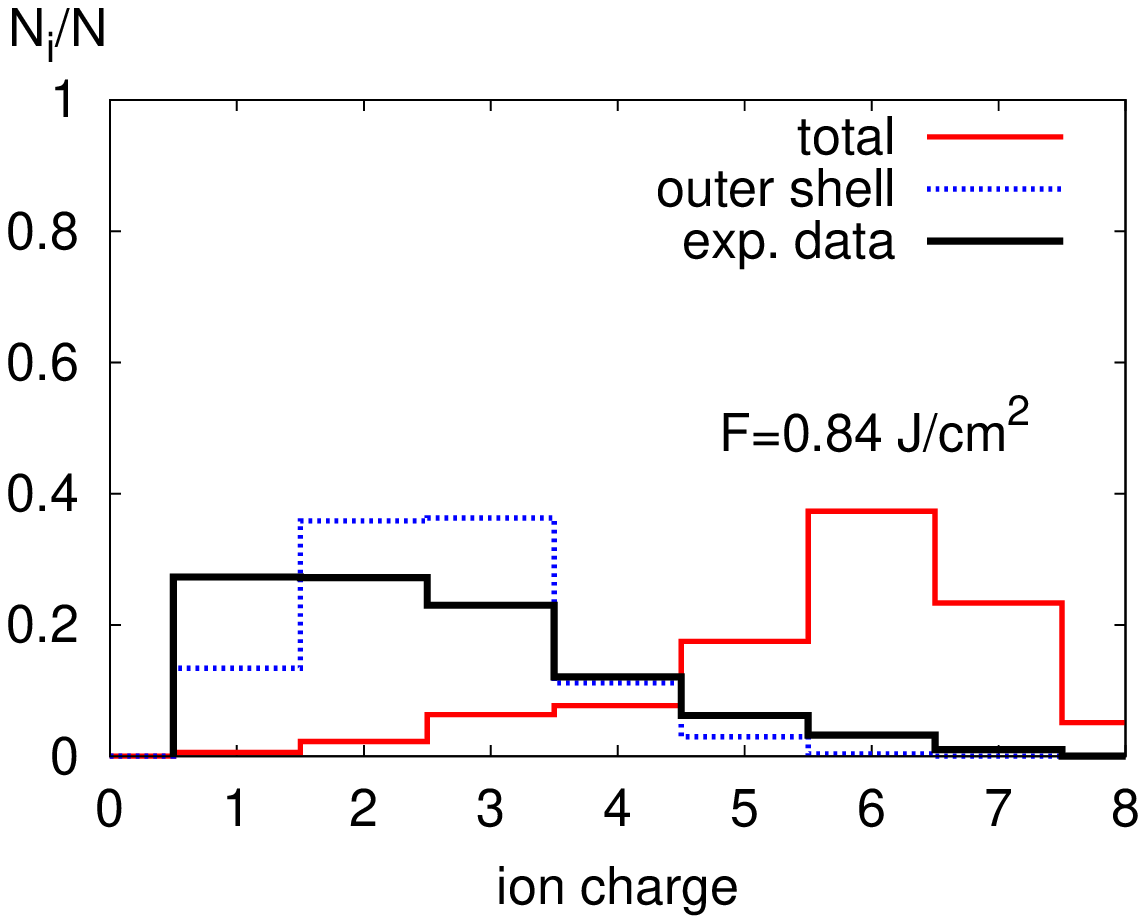}
d)\epsfig{width=6cm, file=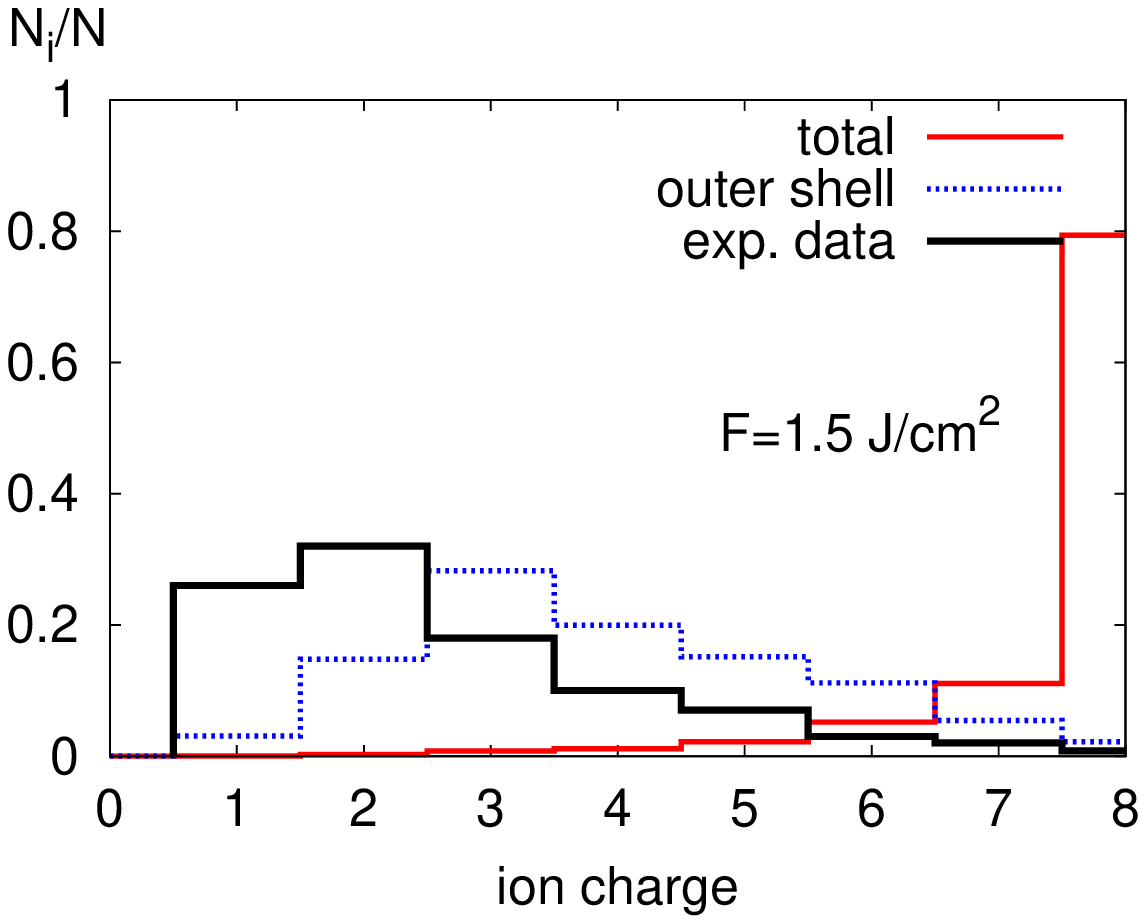} }
\vspace{0.5cm}

\caption{{\it Results with the enhanced IB rate and the atomic potentials of isolated atoms:} Ion fractions within the irradiated xenon cluster at the end of ionization phase calculated within the whole cluster and within the outer shell. They are compared to the experimental data. In each case a)-d) these clusters were irradiated with pulses of different intensities 
($\leq 10^{14}$ W/cm$^2$) and lengths ($\leq 50$ fs) but of a fixed flux. The results obtained were then averaged over the number of pulses. Irradiation at four different radiation fluxes: a) $F=0.05$ J/cm$^2$, b) $F=0.3$ J/cm$^2$, c) $F=0.84$ J/cm$^2$, and d) $F=1.5$ J/cm$^2$ was considered.}
\label{chargen}
\end{figure}

\begin{figure}
\vspace*{0.5cm}
\centerline{\epsfig{width=7cm, file=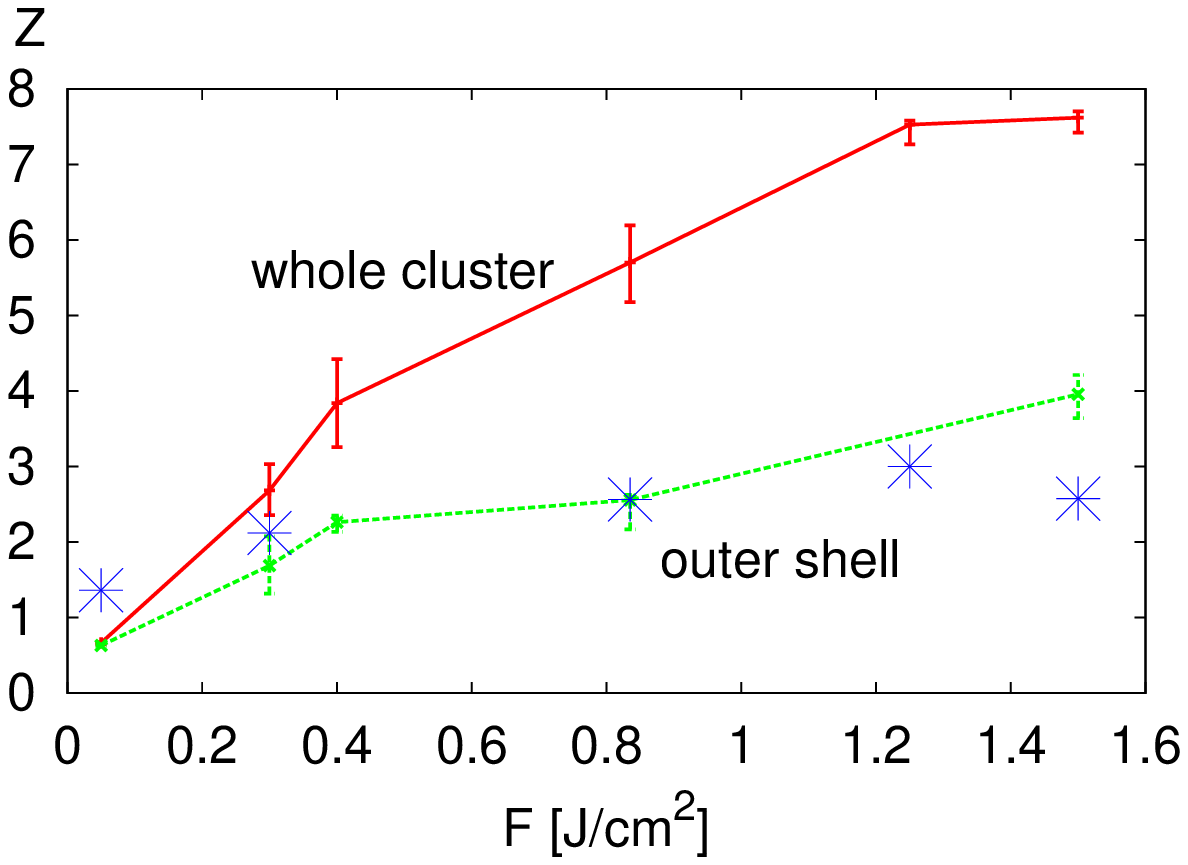}}
\caption{{\it Results with the enhanced IB rate and the atomic potentials of isolated atoms:} Average charge, $Z$, created within the whole irradiated cluster (red errorbars) and within the outer shell (green errorbars) as a function of the time-integrated radiation flux, $F$. These estimates were obtained with pulses of different intensities and lengths but of the fixed flux, and then averaged over the number of pulses. Errorbars denote maximal errors. Experimental data are plotted with stars.}
\label{ladn}
\end{figure}

\begin{figure}
\vspace*{0.5cm}
\centerline{\epsfig{width=7cm, file=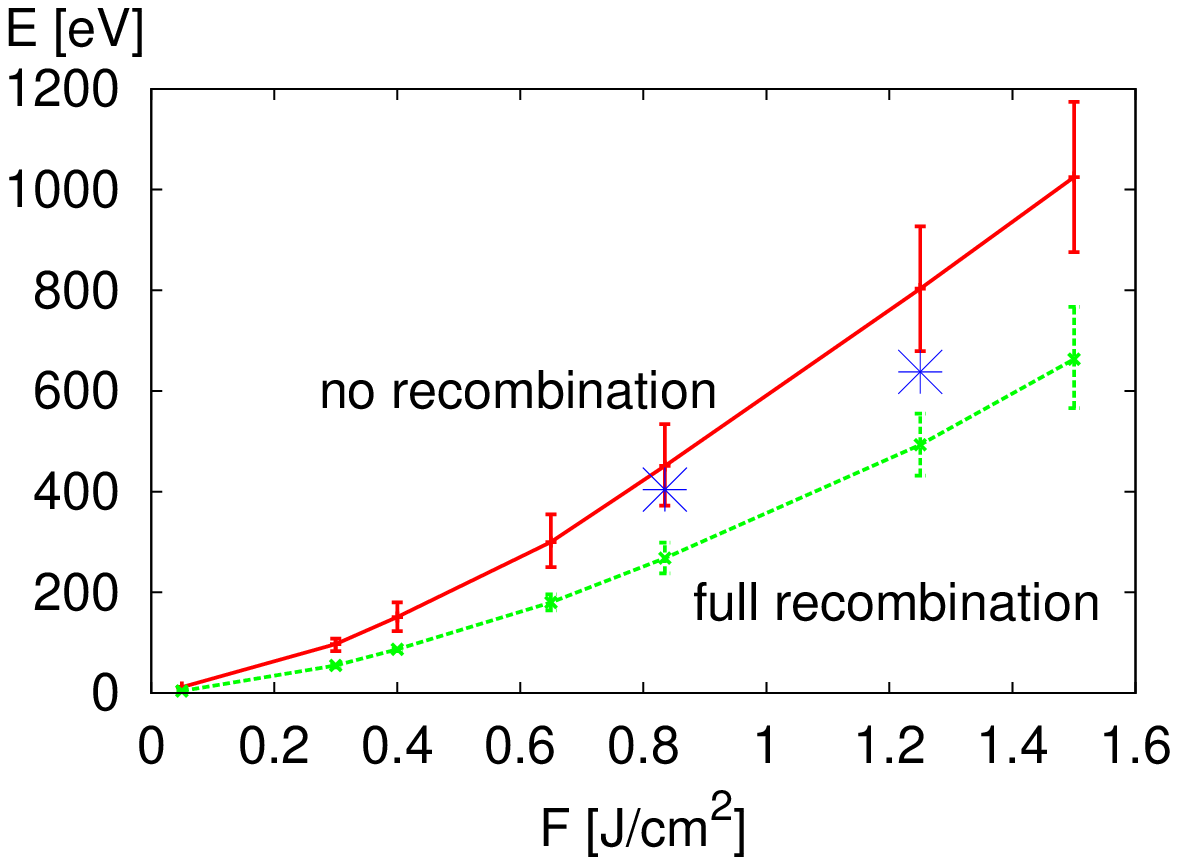}}
\caption{{\it Results with the enhanced IB rate and the atomic potentials of isolated atoms:} Average energy absorbed per atom, $E$, within the irradiated cluster as a function of the time-integrated radiation flux, $F$. Upper
(red errorbars)  and lower (green errorbars) limits for the absorbed energies are estimated within our model. These estimates were obtained with pulses of different intensities and lengths but of the fixed flux, and then averaged over the number of pulses. Errorbars denote maximal errors. Experimental data are plotted with stars.}
\label{energyn}
\end{figure}

Finally, we show the results obtained with the enhanced IB rate and the atomic potentials of isolated atoms/ions. Figs.\ \ref{chargen}, \ref{ladn} and \ref{energyn} show the plots of ion fractions, the average charge and the average absorbed energy. As expected, the estimates obtained are lower than in the previous case in which ionization was faciliated by lowering the ionization potentials. However, the differences are not large, e.g. at the highest flux, $F=1.5$ J/cm$^2$ the Xe$^{+8}$ ion fraction obtained within the whole cluster is: i) $0.85$, when shifts of atomic energy levels are included, and ii) $0.80$ in case of isolated atomic potentials. The total energy absorbed within the whole cluster differs by $\sim 20$ \% at the highest flux. The average charges differ by\\ $\leq$ 10 \% at most.  


\section{Summary}


We performed simulations of ionization dynamics within the $Xe_{2500}$ clusters irradiated with flat VUV pulses of intensity $\leq 10^{14}$ W/cm$^2$ and duration, $\leq 50$ fs. Our model includes the following interactions: photoionizations, collisional ionizations, elastic scatterings of electrons on ions, inverse bremsstrahlung heating, electrostatic interactions of charges, interactions of charges with laser field, shifts of energy levels within atomic potentials due to the plasma environment, and shielded electron-electron interactions.  Limitations and possible improvements of the model are discussed
in the Appendix. 

Within the theoretical framework defined above we studied the impact of various IB rates and the effect of the plasma environment on the overall ionization dynamics. The results obtained were compared to experimental data. We arrived at the following conclusions: 
 
i) all physical mechanisms that were included into the model contributed to the ionization dynamics. The total ionization rate within the sample was affected at most by the heating rate applied, then less strongly by the charge interactions (also the shielded electron-electron interactions) and the plasma environment effects. 
 
ii) the heating rate estimated with Coulomb atomic potentials \cite{krain1,krain} was too low to enable sequential electron impact ionizations leading to the production of charges higher than $+2$. Our analysis included the shifts of the ionization potentials due to the electron screening and to the vicinity of other ions.

iii) high charges up to $+8$ were created with the enhanced IB rate that was estimated with an effective atomic potential \cite{santra}. These high charge states were also observed, when shifts of ionization potentials due to the plasma environment were neglected, i.e. when atomic potentials were approximated with those of isolated atoms and ions. In both cases the total distribution of ion charges obtained with the enhanced IB rate overestimated that one obtained with the experimental data. This effect was especially pronounced at high energy fluxes. 

In analogy to Ref.\ \cite{siedschlag}, the charge distribution within the cluster observed at the end of the ionization phase was inhomogeneous. Cluster consisted of the neutral ion-electron core of the net charge equal to $0$ and of the positively charged outer shell of ions. The highest ion charge was concentrated within the core. The ions of lowest charges could be found only within the surface layer of the cluster (Fig. \ref{out}). The distribution of ions within this surface layer followed the experimental data recorded by the TOF detector. Therefore we expect that the recombination of the core during the expansion phase (not considered here) should significantly reduce the ion charge within the core. This hypothesis should be quantitatively verified with an expansion code.
With the present model we could also obtain the upper and the lower limits of the average energy absorbed per atom. These limits were compared to the available experimental data on the average ejection energy per atom. The experimental data laid within the model estimates.

As we showed above, various processes influence the dynamics of samples irradiated by VUV photons. As there are no full ab initio calculations within the strongly coupled systems, the estimated contribution of these processes can be model-dependent. Dedicated experiments could be helpful in sorting out the relevant models. Experimental estimates of the electron temperature within the irradiated clusters at the end of the ionization phase could verify the  theoretical estimates for the electron heating rate. It is expected that such estimates of the electron temperature could be obtained with cluster experiments similar to the recent holographic experiment that measured the temperature-dependent expansion rate of the irradiated polystyrene spheres \cite{holog}.


\section*{Acknowledgements}

Beata Ziaja is grateful to Cornelia Deiss, Wojciech Rozmus, Robin Santra and Abraham Sz\"oke for illuminating comments. Thomas M\"oller thanks the colleagues from Dresden, especially Jan Michael Rost and Ulf Saalmann, as well as Joshua
Jortner (Tel Aviv) for fruitful discussions. 
This research was supported by the German Bundesministerium f\"ur Bildung und Forschung with grants No.\ 05 KS4 KTC/1 and No.\ 05 KS7 KT1. 


\section{Appendix}


Below we discuss in detail the limitations of our model and propose some improvements.

\noindent
{\bf Single particle densities evolved.}\\   
The applicability of Boltzmann equations is limited to the classical systems which fulfill the assumptions of molecular chaos and two-body collisions. These assumptions are usually justified by a presence of short range forces \cite{shark,rbuick1}. The single particle density function obtained with Boltzmann equations does not contain any information on the three-body and higher correlations. If the higher order correlations are important, a more fundamental Liouville equation for the N-particle density function should be applied \cite{shark}. The Liouville equation reduces to the collisionless Vlasov equation \cite{shark} in case of an uncorrelated system. Fokker-Planck equation \cite{shark} can be derived as a limiting form of the Liouville equation for long-range forces(e.\ g.\ Coulomb forces). It was shown in Ref. \cite{shark} that a correct description of many body Coulomb interactions of plasma electrons and ions as that obtained with the dedicated Fokker-Planck equations can be also obtained with the two-body Boltzmann collision term, assuming the Debye cutoff in the Rutherford scattering cross section. This simplification does not apply to the electron-electron interactions, where the interacting charged particles have identical masses, and the momentum transfer during their collisions cannot be neglected. Therefore we included the respective Fokker-Planck term describing
the shielded electron-electron interactions into our equations. 

\noindent
{\bf Classical evolution.}\\
We describe the evolution of the irradiated samples, using the classical particle densities. After first photoionizations the electron-ion system is dense and strongly coupled. The classical description is then only approximate.  However, as the energy gain by electrons during heating processes is efficient, the system of initially cold electrons enters the classical regime early in the exposure. Classical description is then justified. 

\noindent
{\bf Expansion phase.}\\
Our Boltzmann solver can also follow the expansion phase. However, it becomes computationally inefficient at entering this long semi-equilibrium  evolution phase, as it has still to maintain full stability conditions in both velocity and real space that restrict time steps. On the other hand, there is no need to use the full kinetic equation to follow the semi-equilibrium evolution. At this stage Boltzmann equation can be conveniently replaced by its hydrodynamic limit. Therefore we use Boltzmann solver only to follow the non-equilibrium phase and we stop the evolution of the sample at entering the expansion phase. Simulation of the expansion phase is planned for the forthcoming papers. The three-body recombination rate will also be included into this simulation.

%




\end{document}